\documentclass[aps,twocolumn,pra,amssymb, amsmath,superscriptaddress]{revtex4-2}
\usepackage{dcolumn}
\usepackage{bm}
\usepackage[utf8]{inputenc}
\usepackage[T1]{fontenc}
\usepackage[english]{babel}  
\usepackage{euscript}
\usepackage{amssymb,amsmath,amsfonts,amsthm,bm}
\usepackage{mathtools}
\usepackage{verbatim}
\usepackage{enumerate}
\usepackage{tensor}
\usepackage{fancyhdr}
\usepackage{graphicx}
\graphicspath{{pics/}}
\usepackage[colorlinks=true,linkcolor=blue,citecolor=blue,urlcolor=blue]{hyperref}
\usepackage{bbm}
\usepackage{booktabs}

\usepackage{lineno}
\usepackage{subfigure}
\usepackage[percent]{overpic}
\usepackage{dsfont}
\usepackage{physics}

\theoremstyle{plain}

\theoremstyle{definition}

\usepackage{color}
\definecolor{dred}{rgb}{.8,0.2,.2}
\definecolor{ddred}{rgb}{.8,0.5,.5}
\definecolor{dblue}{rgb}{.2,0.2,.8}
\definecolor{dgreen}{rgb}{.2,0.5,.2}

\newcommand{\expect}[1]{\mbox{$\langle #1 \rangle$}}

\newcommand{\hilb}[1]{{\mathcal{#1}}} 

\newcommand{\subt}[2]{#1_{\mathrm{#2}}}      
\newcommand{\supoper}[1]{\hat{\mathcal{#1}}} 
\newcommand{\ketHS}[1]{\mbox{$|#1 \rangle\rangle$}}
\newcommand{\braHS}[1]{\mbox{$\langle\langle #1|$}}
\newcommand{\ketbraHS}[3]{\mbox{$|#1\rangle\rangle_{#3}\langle\langle #2|$}}
\newcommand{\braketHS}[3]{\mbox{$ \langle\langle{#1}| #2\rangle\rangle_{#3}$}}

\newcommand{\kett}[1]{\left.\ket{#1}\right\rangle}

\newcommand{\phii}[0]{\hat{\Phi}}
\newcommand{\mm}[0]{\hat{\mathcal{M}}}
\newcommand{\uu}[0]{\hat{\mathcal{U}}}
\newcommand{\h}[0]{{\mathcal{H}}}

\begin{document}
\preprint{APS/123-QED}

\title{How coherence measurements of a qubit steer its quantum environment}
\author{Chu-Dan Qiu}\thanks{These authors contributed equally to this work}
\affiliation{State Key Laboratory of Superlattices and Microstructures, Institute of Semiconductors, Chinese Academy of Sciences, Beijing, 100083, China}
\affiliation{Center of Materials Science and Opto-Electronic Technology, University of Chinese Academy of Sciences, Beijing 100049, China}
\author{Yuan-De Jin}\thanks{These authors contributed equally to this work}
\affiliation{Department of Applied Physics, University of Science and Technology Beijing, Beijing 100083, China}
\author{Jun-Xiang Zhang}
\affiliation{Beijing National Laboratory for Condensed Matter Physics and Institute of Physics, Chinese Academy of Sciences, Beijing 100190, China}
\author{Gang-Qin Liu}
\affiliation{Beijing National Laboratory for Condensed Matter Physics and Institute of Physics, Chinese Academy of Sciences, Beijing 100190, China}
\affiliation{Songshan Lake Materials Laboratory, Dongguan, Guangdong 523808, China}
\author{Wen-Long Ma}
\email{wenlongma@semi.ac.cn}
\affiliation{State Key Laboratory of Superlattices and Microstructures, Institute of Semiconductors, Chinese Academy of Sciences, Beijing, 100083, China}
\affiliation{Center of Materials Science and Opto-Electronic Technology, University of Chinese Academy of Sciences, Beijing 100049, China}

\date{\today}

\begin{abstract}
A qubit suffers decoherence when coupled to a classical or quantum environment, and characterizing the qubit decoherence process is vital in quantum technologies. Repetitive Ramsey interferometry measurements (RIMs) are often used to measure qubit coherence, assuming that the environment remains unaffected after each measurement and the outcomes of all measurements are independent and identically distributed (i.i.d.). While this assumption is valid for a classical environment, it may not hold for a quantum environment due to the non-negligible backaction from qubit to environment, especially when the environment has a memory time much longer than the duration of each RIM cycle. Here we present a general theoretical framework to incorporate the measurement backaction from qubit to environment in sequential RIMs. We show that a RIM of a qubit induces a quantum channel on the quantum environment, and sequential RIMs gradually steer the quantum environment to the fixed points of the channel. For the first time, we reveal three distinct environment steering effects---polarization, depolarization and metastable polarization, depending on the commutativity of the noise operator $B$ and the free environment Hamiltonian $H_e$: (1) if $B$ commutes with $H_e$, i.e., $[B,H_e]=0$, the quantum environment is gradually polarized to different eigenstates of $B$ as the number $m$ of repetitive RIMs increases; (2) When $[B,H_e]\neq 0$, the quantum environment is gradually depolarized to a maximally mixed state of its whole Hilbert space or a Hilbert subspace; (3) When $[B,H_e]\neq 0$ but one of $H_e$ and $B$ is a small perturbation on the other, metastable polarization can happen, such that the quantum environment is first polarized for a finite range of $m$ but becomes gradually depolarized as $m$ increases further. The environment steering also makes the measurement statistics of sequential RIMs develop non-i.i.d. features, such that the measurement result distribution can display multiple peaks for a small quantum environment, corresponding to different fixed points of the quantum channel. Realistic examples of central spin models are presented to demonstrate the measurement statistics and environment steering effects. Our work not only elucidates the measurement backaction and statistics of repetitive qubit coherence measurements, but is also useful for designing protocols to engineer the state or dynamics of a quantum environment with a qubit ancilla.
\end{abstract}


\maketitle


\section{Introduction}
Quantum coherence gives rise to a series of non-classical phenomena such as interference and entanglement and lies at the heart of quantum information processing \cite{Streltsov2017}. Realistic physical qubits often suffer loss of coherence (or decoherence) due to coupling to a classical or quantum environment. The environment is called classical if it only imparts random phases to superposition states of the qubit but suffers no backaction from the qubit \cite{Anderson1954,Kubo1954,Cywi2008,Sakuldee2020,Yang2017,Szankowski2017}, and in some cases its effect on the qubit can be well approximated as classical stochastic noise (e.g., Gaussian noise) \cite{Witzel2014,Ma2015,Sakuldee2019}. However, generally the qubit and the environment should be regarded as a composite quantum system, and the qubit has non-negligible backaction on the dynamics of quantum environment, through either the qubit-environment coupling or the measurement and control processes \cite{Deng2006,Zhao2011,Huang2011,Reinhard2012}.

In quantum technologies, it is crucial to characterize the qubit decoherence process by directly measuring the evolution of qubit coherence, either to characterize the quality of physical qubits in quantum computing \cite{Schlosshauer2007,Yang2017} or to reveal useful information about the environments in quantum sensing \cite{Degen2017}. Qubit coherence is conventionally measured by repetitive Ramsey interferometry measurements (RIMs) \cite{Ramsey1950}, often with an underlying assumption that the environment remains unaffected after each RIM and the outcomes of all RIMs are independent and identically distributed (i.i.d.). Then a satisfactory signal-to-noise ratio can be achieved with a relatively large number of repetitions of such RIMs \cite{Jiang2009,Cramer2016}. While this assumption is often valid for a classical environment, it may not hold for a general quantum environment \cite{Beaudoin2018,szankowski2020,Kelly2023}.

Basically a single cycle of RIM slightly alters the state of the quantum environment by inducing a quantum operation on it. If the duration of each RIM cycle is much shorter than the relaxation times of the environment, with repetitive RIMs, the environment can be gradually steered to states quite different from its initial one, and such a steering effect also makes the qubit measurement statistics develop some non-i.i.d. features \cite{Ma2018,Wudarski2023}. Recent experiments in solid state systems have reported signatures of the backaction of qubit measurements on the quantum environment \cite{Cui2012,Hatridge2013,Jerger2023}, and in other experiments such backaction has been utilized to purify a spin bath \cite{Greiner2017,Rao2019,Mkadzik2020}. However, there is still a lack of a general theoretical framework to address the interesting general problems: how repetitive coherence measurements of a qubit steer its quantum environment? What does the measurement statistics look like when considering the environment steering effect?

In this paper, we develop a general theoretical framework to systematically account for the steering effect of sequential qubit coherence measurements on a quantum environment. The essential idea is to model the repetitive RIMs as repetitive quantum channels on the quantum environment. We find that the environment is gradually steered to the fixed points of the quantum channel. More precisely, when noise operator commutes with the environment Hamiltonian, the fixed points of the channel often include rank-one projectors and hence the environment is polarized into the corresponding pure states. Otherwise, the fixed points of the channel include maximally mixed states of the whole Hilbert space or Hilbert subspaces of the environment, so the environment is depolarized to such mixed states if it is initially in a pure state. For both cases, the qubit measurement statistics can display multiple distribution peaks for a small quantum environment, with each peak corresponding to a fixed point of the channel. For realistic scenarios, the noise operator may not commute with environment Hamiltonian, but is only a small perturbation on the environment Hamiltonian. In this case, the environment can exhibit a metastable polarization for a finite range of repetition numbers of RIMs, before relaxing to a depolarized state in the asymptotic limit. As examples, we illustrate the environment steering effects for a central spin embedded in a small quantum spin bath under both RIM sequences and dynamical decoupling (DD) sequences, and elucidate the fine structures of measurement statistics for spin coherence measurements.

We add a comment about the term ``steering'' in this paper. Historically, Schr\"{o}dinger first realized that for a bipartite composite quantum system, one party can influence the wave function of the other party by performing suitable measurements \cite{schrodinger1929}. We use the term ``steering'' in this broader sense \cite{roy2020}, which should be distinguished from the narrower one to denote the impossibility to describe the conditional states at one party by a local hidden state model \cite{Reid2009,Uola2020}.

The paper is organized as follows. In Sec. \ref{basics}, we first introduce the basics of qubit coherence measurement, and then motivate and summarize the main results of the paper. By examining the fixed points of different quantum channels that depend on commutativity between noise operator and environment Hamiltonian, we reveal three distinct environment steering effects---polarization, depolarization and metastable polarization---in Sec. \ref{effect}, which are manifested in the qubit measurement statistics. To elaborate the concepts formalized in previous sections, based on a central spin model in Sec. \ref{example}, we provide several examples with numerical simulations and show that different environment steering effects can be observed in various practical settings.

\section{Preliminaries and main results}

\begin{figure*}[htbp]
    \centering
    \includegraphics[width=2\columnwidth]{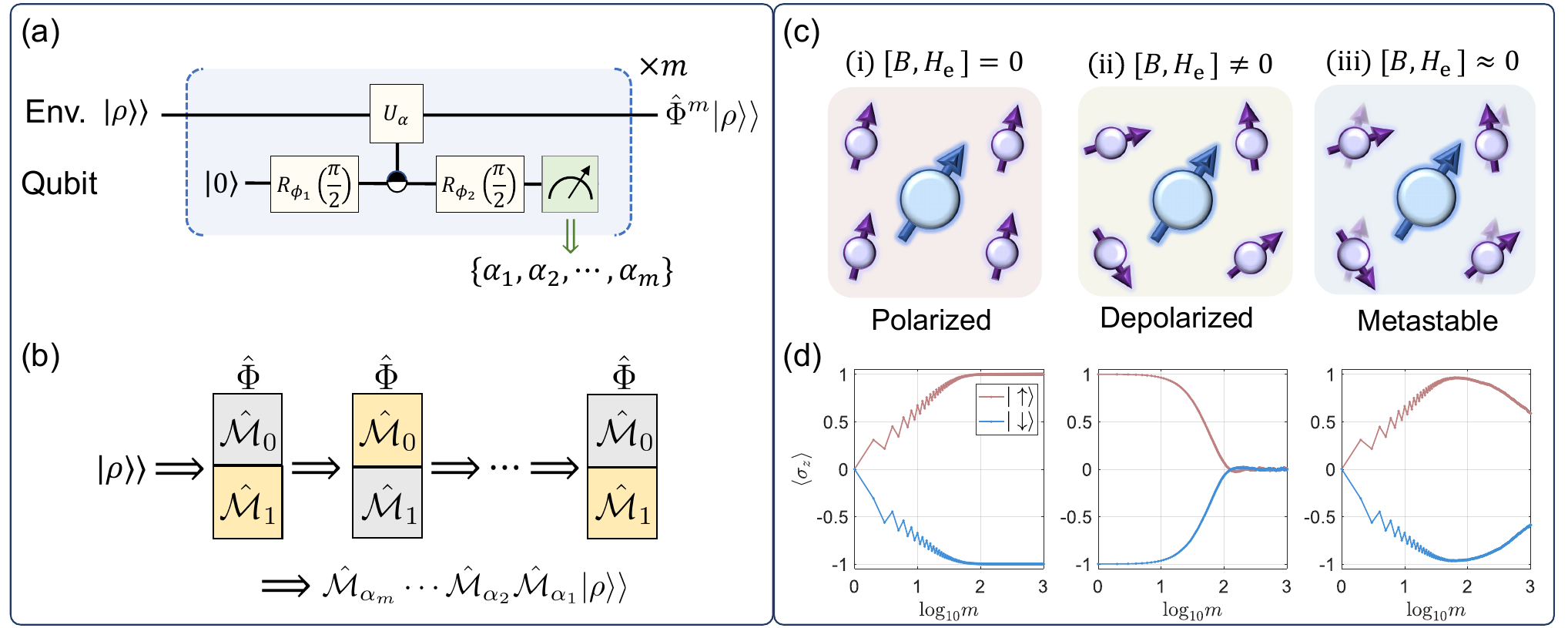}
    \caption{Schematic of sequential RIMs of a qubit and their steering effects on a quantum environment. (a) Environment dynamics under sequential RIMs of a qubit is modeled as sequential applications of a quantum channel $\hat{\Phi}$. In a RIM sequence, the qubit is first initialized into state $\ket{0}_q$ and then undergoes a free evolution sandwiched between two qubit rotations, where $R_{\phi}(\theta)=e^{-i (\cos\phi\sigma_q^x+\sin \phi\sigma_q^y)\theta /2}$ is the qubit rotation operator and $U_{\alpha}=e^{-i[(-1)^{\alpha}B+H_e]t}$ is a unitary operator of the environment conditioned on the qubit state $\ket{\alpha}_q$ ($\alpha=0,1$), and is finally projectively measured with outcome $\alpha_n$ for the $n$th measurement ($1\leq n\leq m$). For an initial state $\kett\rho$ of the environment, its final state becomes $\hat{\Phi}^m|\rho\rangle\rangle$ after repeating RIMs for $m$ times. (b) Sequential channels $\hat{\Phi}^m$ can be decomposed as a summation of all stochastic trajectories, each of which is defined by a sequence of measurement outcomes $\{\alpha_1,\cdots,\alpha_m\}$. Here we only show a typical one. (c) Illustration of quantum environment steering via sequential RIMs of a qubit. For a pure-dephasing coupling between a qubit (blue arrows) and its surrounding environment (purple arrows) [Eq. \eqref{Hamilt}], sequential RIMs can result in drastically different evolutions of the environment. When the noise operator $B$ commutes with the environment Hamiltonian $\subt{H}{e}$, i.e., $[B,H_e]=0$, the quantum environment can be polarized to a certain eigenstate of $B$. But if $[B,H_e]\neq 0$, the quantum environment is depolarized to a maximally mixed state of its whole Hilbert space or a Hilbert subspace. For $[B,H_e]\approx 0$, that is, $[B,H_e]\neq0$ but one of $H_e$ and $B$ is a small perturbation on the other, metastable polarization can happen for some finite range of $m$ before the environment finally relaxes to a depolarized state. (d) Monte Carlo simulations for an illustrative model where $B=\sigma_z$ and $H_{e}=\gamma\sigma_x$ with the parameter $\gamma=0,\ 0.1, \ 0.025$ being an indicator of non-commutativity for three effects respectively. For (i) and (iii), the initial environment state is a maximally mixed state $\mathbb{I}/2$, while for (ii) the initial state is a pure state $\ket{\uparrow}$ or $\ket{\downarrow}$ (eigenstates of $\sigma_z$).}
    \label{1Qbath}
\end{figure*}

\label{basics}

\subsection{Basics of qubit coherence measurements}
Let us consider a single qubit embedded in a $d$-dimensional quantum environment with a pure-dephasing coupling as
\begin{equation}
    \label{Hamilt}
    H = \sigma_q^{z} \otimes B+\mathbb{I}_q \otimes H_e,
\end{equation}
where $\sigma_q^i$ is the Pauli-$i$ operator of the qubit ($i=x,y,z$) with $\sigma_q^z=|0\rangle_{q}\langle0|-|1\rangle_{q}\langle1|$, $B$ is a noise operator for the qubit, $H_e$ is the free Hamiltonian of the environment, and $\mathbb{I}_q$ ($\mathbb{I}$) is the identity operator for the qubit (environment). Hereafter we denote the qubit rotation along an axis in equatorial plane as $R_{\phi}(\theta)=e^{-i(\cos\phi \sigma_q^x+\sin \phi \sigma_q^y)\theta/2}$, where $\phi$ denotes the rotation axis and $\theta$ is the rotation angle. Such a rotation can be generated by a strong qubit control Hamiltonian $H_q=\Omega(\cos\phi \sigma_q^x+\sin \phi \sigma_q^y)/2$ with a duration $\theta/\Omega$.

The qubit coherence can be measured by the widely-used RIM sequence, as shown in Fig. \ref{1Qbath}{\color{blue}(a)}. The qubit is first initialized to state $|0\rangle_q$, and then prepared to be in a superposition state $|\psi\rangle_q=(|0\rangle_q-i e^{i\phi_1}|1\rangle_q)/\sqrt{2}$ by a rotation $R_{\phi_1}(\frac{\pi}{2})$. Then the qubit evolves with the pure-dephasing coupling [Eq. (\ref{Hamilt})] for time $t$, undergoes another rotation $R_{\phi_2}(\frac{\pi}{2})$, and is finally projectively measured in the basis $\{|0\rangle_q,|1\rangle_q\}$. The expectation value of $\sigma_q^z$ after a single RIM is
\begin{equation}\label{Coherence}
    \langle \sigma_q^z\rangle ={\rm Tr}_e\left[\sigma_q^z U({\rho}_{q} \otimes \rho) U^{\dagger}\right]=-{\rm Re}\left\{{\rm Tr}[U_1\rho U_0^{\dagger}] e^{i\Delta\phi}\right\},
\end{equation}
where $\rho_q=|\psi\rangle_q\langle \psi|$, $\rho$ is the initial environment state, $U=\left(R_{\phi_2}(\frac{\pi}{2}) \otimes \mathbb{I}\right)\left(\sum_{\alpha=0,1} |\alpha\rangle_{q}\langle\alpha|\otimes U_{\alpha}\right)$ with $U_{\alpha}=e^{-i[(-1)^\alpha B+H_e]t}$, ${\rm Tr}_e\{\cdot\}$ denotes the partial trace over the environment, ${\rm Re}\{\cdot\}$ denotes the real part of a complex number and $\Delta \phi=\phi_1-\phi_2$.

For sequential RIMs with i.i.d. statistics, by measuring $\langle \sigma_q^z\rangle$ for $\Delta\phi=0$ or $\pi/2$, one can obtain the real or imaginary part of qubit coherence $|{\rm Tr}[U_1\rho U_0^{\dagger}]|$ within the standard quantum limit. Repetitive RIMs display i.i.d. statistics only when the initial environment state is the same for each RIM, but this condition is generally not met for a quantum environment with a much longer memory time compared to the duration of RIM sequence. In this case, the environment state will be slightly altered by each RIM, and sequential RIMs can steer the environment into some stable states determined only by the structures of $B$ and $H_e$, which can be quite different from its initial state. Moreover, the measurement statistics can also develop some non-i.i.d. features.

\subsection{Environment steering of sequential qubit coherence measurements}
The environment steering effect of sequential RIMs can be perfectly described by modeling the RIM sequence as a quantum channel on the quantum environment \cite{Ma2023,Jin2024}. A quantum channel is a completely-positive and trace-preserving (CPTP) map, which can be written in the Stinespring representation as \cite{Stinespring1955}
\begin{equation}\label{Stine}
    \Phi(\rho)={\rm Tr}_{q}\left[U({\rho}_{q} \otimes \rho) U^{\dagger}\right],
\end{equation}
where ${\rm Tr}_q\{\cdot\}$ denotes the partial trace over the qubit. This quantum channel can also be transformed to the Kraus representation as \cite{kraus1983}
\begin{equation}\label{Kraus}
    \Phi(\rho)=M_0\rho M_0^{\dagger}+M_1\rho M_1^{\dagger},
\end{equation}
where two Kraus operators $M_0$ and $M_1$ are related to Hamiltonian given in Eq. \eqref{Hamilt} as
\begin{equation}
    \begin{aligned}
    \begin{bmatrix}
        M_0 \\ M_1
    \end{bmatrix}
    &=\frac{1}{2}
    \begin{bmatrix}
        1 \    &-e^{i\Delta \phi} \\
        1 \    &e^{i\Delta \phi}
    \end{bmatrix}
    \begin{bmatrix}
        U_0 \\ U_1
    \end{bmatrix}\\
    &=\frac{1}{2}
    \begin{bmatrix}
        e^{-i(H_e+B)t}-e^{i\Delta \phi}e^{-i(H_e-B)t}\\
        e^{-i(H_e+B)t}+e^{i\Delta \phi}e^{-i(H_e-B)t}
    \end{bmatrix}.
\end{aligned}
\end{equation}
Note that while $\Phi$ is independent of the second rotation $R_{\phi_2}(\frac{\pi}{2})$, $M_0$ and $M_1$ depend on the phase difference $\Delta\phi$ between the rotation axes of $R_{\phi_1}(\frac{\pi}{2})$ and $R_{\phi_2}(\frac{\pi}{2})$.

Sequential RIM sequences correspond to sequential applications of the same quantum channel on the quantum environment. In this case, it is convenient to use the natural representation of quantum channels \cite{watrous2018}, where the channel can be represented by a single $d^2\times d^2$ matrix. A linear operator on a Hilbert space is transformed to a ket in the Hilbert-Schmidt (HS) space $A=\sum_{ij}a_{ij}|i\rangle\langle j|\leftrightarrow|A\rangle\rangle=\sum_{ij}a_{ij}|ij\rangle\rangle$ with $|ij\rangle\rangle=\ket{i}\otimes \ket{j}$, and the inner product in HS space is defined as $\langle\langle A|B\rangle\rangle=\Tr(A^\dagger B)$. Then the superoperator $X(\cdot)Y$ on the Hilbert space is equivalent to a linear operator $X\otimes Y^{T}$ on the HS space, so the channel $\Phi$ can be naturally represented as
\begin{equation}
  \phii=\mm_0+\mm_1=(\uu_0+\uu_1)/2,\label{RIMKraus}
\end{equation}
where $\hat{\mathcal{M}}_{\alpha}=M_{\alpha}\otimes M_{\alpha}^*$ and $\mathcal{\hat U_\alpha}=U_\alpha\otimes U_\alpha^*$. Note that we add hats for operators acting on the HS space. With the natural representation, the cumbersome $m$-fold nested superoperators on the Hilbert space $\Phi(\cdots\Phi(\Phi(\cdot)))$ is reduced to the $m$th power of the corresponding operator on the HS space $\hat{\Phi}^m$.

Suppose the natural representation of a quantum channel is diagonalizable, then it can be spectrally decomposed as (see Appendix \ref{decomp} for the general case) \cite{wolf2011url}
\begin{equation}
\begin{aligned}
	    \hat\Phi&=\sum_i \lambda_i |R_i\rangle\rangle \langle\langle L_i|,	
\end{aligned}\label{phidecomp}
\end{equation}
where $\{|R_i\rangle\rangle, |L_i\rangle\rangle\}$ is a set of complete biorthonormal bases, i.e., $\braketHS{L_i}{R_j}{}=\delta_{ij}$ with $\delta_{ij}$ being the Kronecker delta. The eigenvalues $\{\lambda_i\}$ of a quantum channel are all located within a unit disk of the complex plane \cite{kraus1983}. The eigenvectors with eigenvalue 1 are called \textit{fixed points} \cite{arias2002,Kribs2003,Burgarth2013} denoted as $\kett{\rho_{\rm fix}^i}$ (every quantum channel has at least one fixed point), those with eigenvalue $e^{i\varphi}$ ($\varphi\neq 0$) are \textit{rotating points} \cite{Albert2019}, and those with $\abs{\lambda_i}<1$ are \textit{decaying points}. We will show below that the fixed points of the channel in Eq. (\ref{RIMKraus}) depend on the commutativity of $B$ and $H_e$ (see Appendix \ref{Proposition}): (1) if $[B,H_e]=0$, the fixed points are spanned by a set of rank-one projections $\{|j\rangle\langle j|\}_{j=1}^d$; (2) if $[B,H_e]\neq 0$, the fixed points are spanned by a set of projection operators $\{\Pi_j\}_{j=1}^r$ ($r<d$), satisfying $\sum_{j=1}^r\Pi_j=\mathbb{I}$.


With the diagonalized form of a quantum channel, it becomes easier for us to investigate its asymptotic behavior. Evidently, as the number of measurements $m$ increases, the decaying points tend to vanish as the number of repetitions $m$ increases. While the states in the subspace spanned by the fixed points and rotating points, which is called asymptotic subspace (also known as peripheral or attractor subspace), either remain unchanged or only acquire a phase during repetition. Therefore, the asymptotic limit of sequential quantum channels is a projection to the asymptotic subspace. If $\hat\Phi$ has no rotating points, the asymptotic behavior of sequential quantum channels is determined solely by its fixed points as \cite{Albert2019}, (see Appendix \ref{decomp}):
\begin{equation}
    \label{QChannelAsympt_inf}
    \lim_{m \rightarrow \infty}\subt{\hat{\Phi}}{}^m =\sum_{j=1}^J \hat{\mathcal{P}}_j= \sum_{j=1}^J \ketbraHS{\rho^{j}_{\mathrm{fix}}}{P^{j}_{\mathrm{fix}}}{},
\end{equation}
where $\{\rho^{j}_{\mathrm{fix}}\}_{j=1}^J$ is a set of bases for the HS subspace spanned by the fixed points, which contains positive operators with unit trace [Tr$(\rho_{\rm fix}^j)$=1 for any $j$] and orthogonal supports ($\rho_{\rm fix}^i\rho_{\rm fix}^j=0$ if $i\neq j$), and $P^{j}_{\mathrm{fix}}$ are a set of observables satisfying $\braketHS{P^{i}_{\mathrm{fix}}}{\rho^{j}_{\mathrm{fix}}}{}=\delta_{ij}$. For the channel induced by a RIM, we have: (1) if $[B,H_e]=0$, $\rho_{\rm fix}^j=P^{j}_{\mathrm{fix}}=|j\rangle\langle j|$ ($1\leq j\leq d$); (2) if $[B,H_e]\neq0$, $\rho_{\rm fix}^j=P^{j}_{\mathrm{fix}}/d_j=\Pi_j/d_j$ with $d_j$ denoting the rank of $\Pi_j$ ($1\leq j\leq r$).

\subsection{Statistics of sequential qubit coherence measurements}
With the environment steering effect, the statistics of sequential qubit coherence measurements can develop some non-i.i.d. features. For a single RIM cycle, the probability to obtain the measurement result $\alpha\in\{0,1\}$ is
\begin{equation}\label{mprob}
    p_{\alpha}=\Tr(M_{\alpha} \rho M_{\alpha}^{\dagger})=\braHS{\mathbb{I}} \supoper{M}_\alpha \ketHS{\rho},
\end{equation}
from which one can obtain the qubit coherence as $\langle \sigma_q^z\rangle=p_0-p_1=\langle\langle\mathbb{I}|(\supoper{M}_0-\supoper{M}_1)|\rho\rangle\rangle$. The probability distribution for result 0 and 1 is denoted as $F'=(p_0,p_1)$. For $m$ sequential RIMs, we get a sequence of binary numbers $\{\alpha_1, \cdots, \alpha_m\}$ with $\alpha_n\in\{0,1\}$ for any $n\in[1,m]$, and we record the frequency of result $\{0, 1\}$ as $F=(f_0,f_1)=(\frac{m_0}{m},\frac{m_1}{m})$, where $m_0(m_1)$ are the numbers of occurrences of $0(1)$ respectively satisfying $m_0+m_1=m$. A sequence of measurement outcomes defines a quantum trajectory. In practice, we often focus on the average of the $m$ measurement results, i.e., $\bar{\alpha}=\frac{1}{m}\sum_{n=1}^m \alpha_n$, which coincides with $f_1$ in our case. If the outcomes of different RIMs are i.i.d., $F$ obeys a binomial distribution:
\begin{equation}\label{Fc}
    p(F)=\frac{m!}{(mf_0)!(mf_1)!}p_0^{mf_0}p_1^{mf_1}.
\end{equation}
Hence, the expectation of measurement average can be readily obtained as $\expect{f_1}=\sum_F f_1  p(F) = p_1$, where $p_1$ is determined by Eq. \eqref{mprob}. According to the central limit theorem, a binomial distribution can be approximated by a Gaussian distribution as $m \rightarrow \infty$. In other words, the statistics of qubit coherence measurements concentrate around the probability distribution $F'=(p_0,p_1)$ for large $m$.

However, when we consider the backaction of qubit coherence measurements on the quantum environment, the i.i.d. assumption does not hold. The probability to get a specific sequence of measurement results $\{\alpha_1,\cdots,\alpha_m\}$ is
\begin{equation}\label{pniid}
    p(\alpha_1,\cdots,\alpha_m|\rho)=\braHS{\mathbb{I}} \supoper{M}_{\alpha_m} \cdots \supoper{M}_{\alpha_1} \ketHS{\rho},
\end{equation}
Then what is the distribution of the measured frequency $F$? Is it still concentrated around $F'$? We will show in next section that this is generally not the case.

More specifically, based on decomposing a channel in HS space into two orthogonal parts as $\hat{\Phi} = \hat{\mathcal{P}} + \subt{\hat{\Phi}}{D}$, where the first term is the projection into fixed point space [Eq. \eqref{QChannelAsympt_inf}], we show that the expectation of measurement average takes the form (see Appendix \ref{exp0})
\begin{equation}
    \label{stat}
    \begin{aligned}
        \expect{f_1} &= \sum_{\alpha_1} \cdots \sum_{\alpha_m} f_1 p(\alpha_1,\cdots,\alpha_m|\rho)\\
        &=\sum_{j=1}^J c_j \expect{f_{1j}}_{*}+\frac{1}{m} \langle\langle \mathbb{I} |\supoper{M}_1 \sum_{n=1}^{m}\subt{\hat{\Phi}}{D}^{n-1}\supoper{Q} |\rho\rangle\rangle,
    \end{aligned}
\end{equation}
where
\begin{equation}\label{f1fixed}
    \expect{f_{1j}}_{*} =\braketHS{\mathbb{I}}{\supoper{M}_1|\rho^{j}_{\mathrm{fix}}}{}=p(1|\rho^{j}_{\mathrm{fix}})
\end{equation}
is the probability to get result $\alpha=1$ for the fixed point $\rho^{j}_{\mathrm{fix}}$, $c_j=\Tr(P^{j}_{\mathrm{fix}} \rho)$ denotes the length of the projection of $\rho$ to $\rho^{j}_{\mathrm{fix}}$ in the HS space, $\supoper{Q}=\supoper{I}-\hat{\mathcal{P}}$ is the projection to the orthogonal complement of the fixed point space with $\supoper{I}$ being the identity operator in HS space. It can be verified that in the asymptotic limit, the first term contributed by the fixed point space dominates, while the second term becomes insignificant, which is consistent with Eq. \eqref{QChannelAsympt_inf} (see Appendix \ref{qstatisticsLim}). In other words, the distribution of measured frequency can display $J$ peaks (without degeneracy) that center around the expectation determined by each fixed point $\{\expect{f_{1j}}_{*}\}_{j=1}^J$ respectively. Each peak includes trajectories defined by $\{\alpha_1, \cdots ,\alpha_m\}$ with frequency $f_1$ close to $\expect{f_{1j}}_*$, which can steer an arbitrary initial environment state to one of fixed point subspaces \cite{Ma2023} [see Fig. \ref{1Qbath}{\color{blue}(b)} and the following sections for further discussions].

On the other hand, we can further rewrite $\expect{f_{1j}}_{*}$ in Eq. \eqref{f1fixed} as
\begin{equation}
    \expect{f_{1j}}_{*}=\Tr(M_1 \rho^{j}_{\mathrm{fix}} M_1^\dagger)=\frac{1-\expect{\sigma_q^z}_{j*}}{2},
\end{equation}
where $\expect{\sigma_q^z}_{j*}$ is the coherence given in Eq. \eqref{Coherence} and the subscript indicates its dependence on the fixed points of environment $\rho^{j}_{\mathrm{fix}}$. This suggests that only when the environment is initially in one fixed point $\rho^{j}_{\mathrm{fix}}$ or the convex combination of fixed points, i.e.,  $\sum_{j=1}^J a_j\rho^{j}_{\mathrm{fix}}$ with $a_j\geq 0$ and $\sum_{j=1}^J a_j=1$, the coherence obtained from $m$-fold sequential RIMs is equivalent to the one obtained from applying a single RIM on $m$ identical copies of the initial environment state respectively. And the coherence with respect to each fixed point can be extracted from the central location of each peak $\expect{f_{1j}}_{*}$.

In previous studies concerning qubit decoherence, e.g., cluster-correlation expansion (CCE) theory to calculate spin decoherence in a many-body spin bath \cite{yang2008,yang2009,Onizhuk2023bath,onizhuk2024}, the environment is implicitly assumed to remain unaffected during qubit coherence measurements, and thus qubit decoherence can be directly calculated from a single RIM cycle as in Eq. \eqref{Coherence}. The results in this paper show that the calculations with maximally mixed state as the initial environment state are still valid even when considering the environment steering effect, since the maximally mixed state can be represented as a convex combination of fixed points. However, this may not the case for an arbitrary initial environment state. So it requires further study to explore the reasonable initial environment states for correct prediction of qubit decoherence in complex quantum environments. We will not address this issue in this paper, but only focus on the steering effects of qubit coherence measurements on a small quantum environment.

\section{Quantum environment steering and non-i.i.d. measurement statistics}\label{effect}

Sequential qubit coherence measurements can show different environment steering effects and measurement statistics, which in turn depend on the structures of the noise operator $B$ and the free environment Hamiltonian $H_e$ [Eq. \eqref{Hamilt}]. We elaborate on three cases [see Fig. \ref{1Qbath}{\color{blue}(c-d)}]: (1) $[B,H_e]= 0$: sequential RIMs polarize the environment to a pure state; (2) $[B,H_e] \neq 0$: sequential RIMs depolarize the environment in the whole Hilbert space or a Hilbert subspace. (3) $[B,H_e]\approx 0$ ($[B,H_e]\neq 0$, but one of $H_e$ and $B$ is a small perturbation on the other): sequential RIMs induce metastable polarization for a finite range of repetitions. In all cases, the measurement statistics can display multiple distribution peaks for a small quantum environment, with each peak corresponding to a fixed point of the channel.

\subsection{Polarization: $[B, {H}_{e}]= 0$}
If $[B, {H}_{e}]= 0$, they can be simultaneously diagonalized as
\begin{equation}\label{VHc}
    B=\sum_{k=1}^s b_k P_k, ~~~~ H_e=\sum_{k=1}^s \varepsilon_k P_k,
\end{equation}
where $b_k$, $\varepsilon_k$ are the $k$th eigenvalues of $B$ and $H_e$ respectively, and $P_k$ is the projection operator to the eigenspace $\mathcal{H}_k$ satisfying $\sum_{k=1}^{s}P_k=\mathbb{I}$ so that the Hilbert space of the environment is decomposed as $\mathcal{H}=\bigoplus_{k=1}^s\mathcal{H}_k$. Then the Kraus operators can be recast as \cite{Ma2023}
\begin{equation}
    \label{KrausDecomp}
    \begin{bmatrix}
        M_0 \\ M_1
    \end{bmatrix}=
    \begin{bmatrix}
        {\tilde{\lambda}}_{01} \  &\cdots \  &{\tilde{\lambda}}_{0s} \\
        {\tilde{\lambda}}_{11} \  &\cdots \  &{\tilde{\lambda}}_{1s}
    \end{bmatrix}
    \begin{bmatrix}
        P_1 \\ \vdots \\ P_s
    \end{bmatrix},
\end{equation}
where $\tilde{\lambda}_{\alpha k}=e^{-i\varepsilon_kt}[e^{-i b_kt}-(-1)^{\alpha}e^{i(\Delta\phi+b_kt)}]/2$ is the $k$th eigenvalue of $M_{\alpha}$. Note that $\tilde{\bm{\lambda}}_k=[\tilde{\lambda}_{0k},\tilde{\lambda}_{1k}]^T$ is a unit column vector in a two-dimensional complex vector space due to $\sum_{\alpha=0,1}M_{\alpha}^\dagger M_{\alpha} =\mathbb{I}$, and $\{\tilde{\bm{\lambda}}_{k}\}_{k=1}^{s}$ is a set of such unit vectors. Then $\supoper{M}_{\alpha}$ and $\hat{\Phi}=\sum_{\alpha}^{} \supoper{M}_{\alpha}$ are all diagonal operators on the HS space,
\begin{equation}\label{QchannelHSc}
    \hat{\Phi}=\sum_{k,l=1}^s \langle\tilde{\bm{\lambda}}_{l},\tilde{\bm{\lambda}}_{k}\rangle P_k \otimes P_l,
\end{equation}
where the eigenvalue of channel $\langle\tilde{\bm{\lambda}}_{l},\tilde{\bm{\lambda}}_{k}\rangle$ is the inner product between $\tilde{\bm{\lambda}}_{l}$ and $\tilde{\bm{\lambda}}_{k}$ in the Euclidean vector space, and $P_k \otimes P_l$ is a projection operator on the HS space. Since $|\langle\tilde{\bm{\lambda}}_{l},\tilde{\bm{\lambda}}_{k}\rangle|\leq\langle\tilde{\bm{\lambda}}_{k},\tilde{\bm{\lambda}}_{k}\rangle\langle\tilde{\bm{\lambda}}_{l},\tilde{\bm{\lambda}}_{l}\rangle=1$ due to the Cauchy-Schwarz inequality, all the eigenvalues of $\hat{\Phi}$ lie within the unit disk of the complex plane.
Then sequential applications of channel $\hat{\Phi}$ produce a projective measurement in the asymptotic limit, as given in Eq. \eqref{QChannelAsympt_inf},
\begin{equation} \label{QChannelCAsympt_inf}
    \lim_{m \rightarrow \infty}{\hat{\Phi}}^m =\sum_{k=1}^s\supoper{P}_k,
\end{equation}
where $\supoper{P}_k=P_k \otimes P_k\leftrightarrow P_k(\cdot)P_k$ corresponds to a projection to $\mathcal{H}_k$. So any state $\rho_k\in \mathcal{H}_k$ is a fixed point, and $\rho^{k}_{\mathrm{fix}}=P^{k}_{\mathrm{fix}}$ in this case (see Appendix \ref{Proposition}).

For measurement statistics, since $[\hat{\mathcal{M}}_{0}, \hat{\mathcal{M}}_{1}]=0$, results similar to Eq. \eqref{stat} can be obtained by expanding $\hat{\Phi}^m=\sum_{F}\hat{\Phi}^m(F)$ according to the binomial theorem,
\begin{equation}
    \hat{\Phi}^m(F)=\frac{m!}{(mf_0)!(mf_1)!}\hat{\mathcal{M}}_0^{mf_0}\hat{\mathcal{M}}_1^{mf_1}.
\end{equation}
In the asymptotic limit, $\hat{\Phi}^m$ can be approximated by its projections to the asymptotic space \cite{Ma2023},
\begin{equation}\label{QChannelC}
    \hat{\Phi}^m(F)\approx \sum_{k=1}^s\supoper{P}_k \hat{\Phi}^m(F) \supoper{P}_k \approx \sum_{k=1}^s e^{-mS(F\|F_k)}\supoper{P}_k,
\end{equation}
where $S(F\|F_k)=\sum_{\alpha=0,1} f_\alpha\ln (f_\alpha/p_{\alpha k})$ is the relative entropy between $F$ and $F_k$ with $F_k=(p_{0k},p_{1k})=(|\tilde{\lambda}_{0k}|^2, |\tilde{\lambda}_{1k}|^2)$. Accordingly, the measurement statistics can be exactly obtained as
\begin{equation}
    p(F) = \braHS{\mathbb{I}} \hat{\Phi}^m(F) \ketHS{\rho} = \sum_{k=1}^s e^{-mS(F\|F_k)} \Tr(P_k \rho).
\end{equation}
Compared to Eq. \eqref{Fc}, the probability can be rewritten as $p(F)=\sum_{k=1}^s p(F|P_k)p(P_k|\rho)$, where $p(P_k|\rho)=\Tr(P_k \rho)$ is the probability to find the environment state in $\mathcal{H}_k$, and $p(F|P_k)\approx e^{-mS(F\|F_k)}$ is the conditional probability to obtain a measured frequency distribution $F$ given the environment state $P_k$.

For large $m$, Eq. (\ref{QChannelC}) can be further approximated by $s$ Gaussians around $F_k$ respectively, since $S(F\|F_k)\approx \sum_{\alpha=0,1} (f_{\alpha}-p_{\alpha k})^2/(2p_{\alpha k})$. Then the expectation value of $f_1$ can be obtained
\begin{equation}\label{statC}
    \langle f_1\rangle=\sum_{F} f_1 p(F)=\sum_{k=1}^s \Tr(P_k \rho)\expect{f_{1k}}_{*},
\end{equation}
where
\begin{equation}\label{f1fixedc}
    \expect{f_{1k}}_{*} =p_{1k}=|\tilde{\lambda}_{1k}|^2=\frac{1}{2}\left[1+\cos(2b_k t+\Delta \phi)\right],
\end{equation}
corresponding to the expectation of $f_1$ for the $k$th Gaussian, in agreement with Eq. \eqref{f1fixed}. Owing to $\rho^{k}_{\mathrm{fix}}=P^{k}_{\mathrm{fix}}=P_k$, Eq. \eqref{statC} is consistent with Eq. \eqref{stat}, where $c_k=\Tr(P_k \rho)=p(P_k|\rho)$. Note that the second term in Eq. \eqref{stat} is absent here since we consider only projections to the asymptotic space in Eq. \eqref{QChannelC}. If the initial state is a fixed point $\rho_k$, the probability of obtaining a sequence of measurement outcomes [Eq. \eqref{pniid}] can be reduced to $p(\alpha_1,\cdots,\alpha_m|\rho_k)= \prod_{n=1}^{m} p({\alpha_n|\rho_k})=\prod_{n=1}^{m} p_{\alpha_n,k}$, as if measurements were performed independently on $m$ identical copies of the initial state ${\rho}_k$. This is similar to i.i.d. measurements in the classical scenario.

Moreover, even if the initial environment state is a mixed state, it is possible to polarize the environment to certain fixed point $\rho_k$ by selecting the trajectories with frequency distribution close to $F_k$. Two distributions around $F_j$ and $F_k$ are well separated if the distance between $F_j$ and $F_k$ is larger than the sum of the respective half widths. If all the binomial distributions are well separated, integration of $\hat{\Phi}^m(F)$ within a small neighborhood around $F_k$ can approximate $\hat{\mathcal{P}}_k$ up to arbitrary small error as $m$ increases \cite{Ma2023}.


\subsection{Depolarization: $[B, H_e] \neq 0$}
If $[B, H_e] \neq 0$, we can simultaneously reduce $B$ and $H_e$ to a block-diagonal form by an appropriate unitary transformation $W$,
\begin{equation}\label{VHq}
    B=W\left(\bigoplus_{j=1}^r B_j\right) W^{\dagger}, ~~~~
    H_e= W\left(\bigoplus_{j=1}^r H_{j}\right) W^{\dagger}.
\end{equation}
This partitions the Hilbert space $\mathcal{H}$ of the quantum environment into the direct sum of $r$ subspaces $\mathcal{H}=\bigoplus_{j=1}^r\mathcal{H}_j$, and $B_j$, $H_j$ are operators acting on the subspace $\mathcal{H}_j$ (here $W$ should be chosen so that $B_j$ and $H_j$ for any $j$ cannot be reduced further to another block-diagonal form). Then $[B_j, H_{j}] \neq 0$ for at least one subspace $\mathcal{H}_j$ with ${\rm dim}(\mathcal{H}_j)\geq2$, while Eq. (\ref{VHc}) can be regarded as a special case of Eq. (\ref{VHq}) with ${\rm dim}(\mathcal{H}_j)=1$ for all $j$.

One can show that the channel ${\Phi}$ is unital, i.e., $\Phi(\mathbb{I})=\mathbb{I}$, so the maximally mixed state $\mathbb{I}/d$ is a fixed point. The full set of fixed points can be obtained by using a theorem in \cite{watrous2018}: $\rho$ is a fixed point of a unital channel if and only if it commutes with every Kraus operator, i.e., $[\rho, M_\alpha]=0$ for any $\alpha$. Then for the noise and environment operators in Eq. (\ref{VHq}), the fixed points of the corresponding channel are spanned by a set of projection operators $\{\Pi_j\}_{j=1}^r$, where $\Pi_j$ is the projection to the subspace $\mathcal{H}_j$ satisfying $\sum_{j=1}^r\Pi_j=\mathbb{I}$ (see Appendix \ref{Proposition}). Then sequential applications of channel $\hat{\Phi}$ produce a depolarizing operation in the asymptotic limit,
\begin{equation}\label{ChannelQAsympt_inf}
    \lim_{m \rightarrow \infty}{\hat{\Phi}}^m
              =\sum_{j=1}^r \ketbraHS{\Pi_j/d_j}{\Pi_j}{}.
\end{equation}

The depolarization can be understood from measurement statistics. In this case, the probability of obtaining a sequence of measurement results described by Eq. \eqref{pniid} is generally non-i.i.d.. Nevertheless, noting that the identity is the left eigenvector of the channel, i.e., $\braHS{\mathbb{I}}\hat{\Phi}=\braHS{\mathbb{I}}$, we can obtain the expectation \cite{Burgarth2015} (see Appendix \ref{exp0})
\begin{equation}
    \expect{f_{1}} =\sum_{j=1}^r c_j\expect{f_{1j}}_{*}+\frac{1}{m} \langle\langle \mathbb{I} |\supoper{M}_1 \sum_{n=1}^{m}\subt{\hat{\Phi}}{D}^{n-1}\supoper{Q} |\rho\rangle\rangle,
\end{equation}
around which the peaks of measured frequency distribution are centered, and $\expect{f_{1j}}_{*}$ is given in Eq. \eqref{f1fixed}. Particularly, for a $d_j$-dimensional subspace where $[B_j, H_{j}] \neq 0$ the fixed point is a maximally mixed state $\mathbb{I}/d_j$, the only one peak concentrates around $\expect{f_{1j}}_{*} = (1-\expect{\sigma_q^z}_{\mathbb{I}})/2$, which is independent of any initial environment state, even if the environment starts from a pure state. As $m$ increases, it can be demonstrated that the first term, representing the contributions from fixed point space, become dominant (see Appendix \ref{qstatisticsLim}). This is also manifested in the narrowing of peak broadening, which we will show in the Sec. \ref{example} and Appendix \ref{var0}.

\subsection{Metastable polarization: $[B, H_e] \approx 0$}
If $[B,{H}_{e}]\approx 0$, that is, $[B,H_e]\neq 0$ but one of $H_e$ and $B$ is a small perturbation on the other, then for the channel [Eq. (\ref{RIMKraus})] with $r$ fixed points, it may have $q-r \ (d\leq q< d^2)$ decaying points with eigenvalues close to one $|\lambda_j| \approx 1$ for $j\in [r+1,\cdots,q]$ (where we sort the eigenvalues in descending order). Then metastable polarization can happen such that the quantum environment is first polarized for a finite range of $m$ \cite{Macieszczak2016,Jin2024}, which is determined by the gap between $\lambda_q$ and $\lambda_{q+1}$, 
 \begin{equation}\label{meta}
    \frac{1}{\big|\ln|\lambda_{q+1}|\big|}\ll m\ll \frac{1}{\big|\ln|\lambda_q|\big|},
\end{equation}
but as $m$ increases further beyond this range, the environment becomes gradually depolarized to maximally mixed states in one of the subspaces $\{\mathcal{H}_j\}_{j=1}^r$, corresponding to the fixed points of the channel.

Accordingly, the dynamics of metastable polarization can be observed from measurement statistics as $m$ changes. Take a $d_j$-dimensional subspace as an example, whose fixed point is the maximally mixed state of the subspace $\Pi_j/d_j$. Within the regime given by Eq. \eqref{meta}, approximate projective measurements give rise to $d_j$ peaks in measured frequency distribution. However, when $m$ increases beyond the metastable regime, the depolarization gradually dominates, while peaks become indistinguishable and finally merge into one peak which corresponds to the only fixed point $\Pi_j/d_j$.

\section{Application to a central spin model}
\label{example}
We illustrate the environment steering effects and statistics of sequential qubit coherence measurements with a central spin model, where an electron spin (as the central qubit) is immersed in a small nuclear spin bath (as the environment) with long coherence time. This model is applicable for many electron-nuclear spin systems. For concreteness, we use parameters of a nitrogen-vacancy (NV) center with a carbon nuclear spin bath for simulations (All simulations in this section contain $2 \times 10^4$ samples with $\Delta \phi=\pi/2$, and the initial state is the maximally mixed state of the whole Hilbert space).

\subsection{Single-spin bath}
Let us first consider the simplest spin bath, namely, a bath that has only one spin-1/2.

\subsubsection{RIM sequence}
\begin{figure}[tp]
    \centering
    \includegraphics[width=\columnwidth]{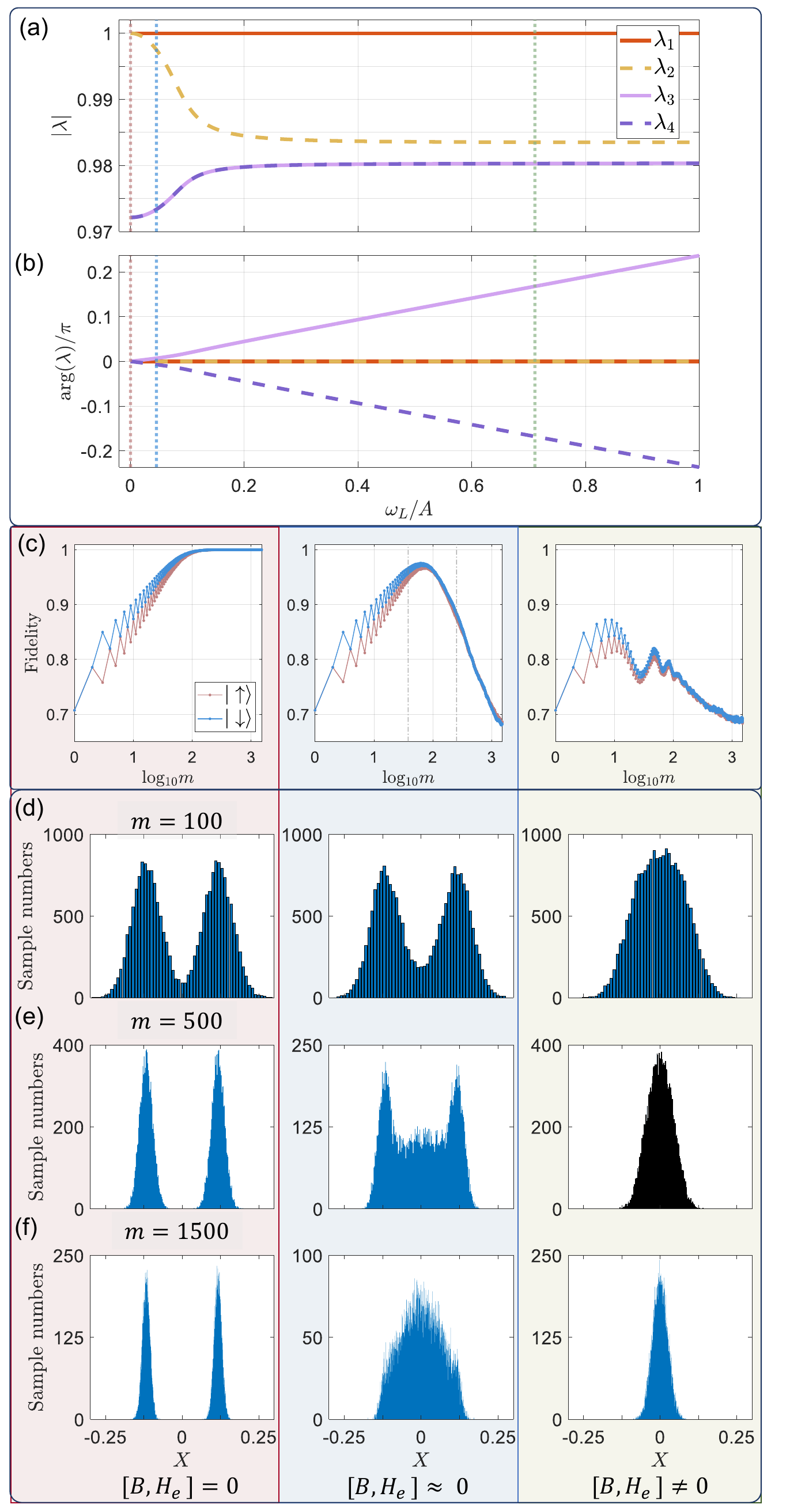}
    \caption{Monte Carlo simulations of a single-spin bath steering via sequential RIMs [Eq. \eqref{singleq}]. (a) Modulus and (b) argument of eigenvalues of the quantum channel when varying external magnetic field. For different Larmor frequencies ($\omega_L/A=0, \ 0.06,\ 0.71$, labeled by red, blue and green respectively), polarization, metastable polarization and depolarization can be observed. For each effect, we show (c) fidelity and (d-f) qubit measurement statistics with different number of repetitions $m$ where $X =(f_{1}-f_{0})/2$. Each peak of measurement statistics corresponds to a fixed point of quantum channel. Fidelity $\mathrm{F}(\rho_{\mathrm{fix}},\rho)=\Tr\left(\sqrt{\sqrt{\rho}\ \rho_{\mathrm{fix}}\ \sqrt{\rho}}\right)$ is obtained by averaging trajectories leading to the same fixed points $\rho_{\mathrm{fix}}$. Here trajectories with $X\in [-0.5,0)$ ($X \in [0,0.5]$) are regarded as projections to $\ket{\uparrow}$ ($\ket{\downarrow}$), which are eigenstates of $B$. The central locations of peaks are determined by $\expect{X}_*$ according to Eq. \eqref{f1fixed}. Particularly, when $[B,H_e]=0$, by tuning free evolution time $t$, the strength of measurement, manifested in $\expect{X_{\pm}}_*=\cos({\pm}At+\Delta \phi)/2$, can be tuned accordingly. Here we use $A=37.7$ kHz, $t=1 \  \mu s$, $\Delta \phi=\frac{\pi}{2}$.}
    \label{2simul1nuclRamsey}
 \end{figure}

For sequential RIMs of a central spin, the noise operator and free Hamiltonian of the bath spin are
\begin{equation}\label{singleq}
  B=\bm{A}\cdot\bm{I}, \quad H_e=\omega_L I_z,
\end{equation}
where $\bm{A}=(A_{x},A_{y},A_{z})$ is a hyperfine vector between central qubit and bath spin with $A=|\bm{A}|$, $\bm{I}=(I_x, I_y, I_z)$ is the spin-1/2 vector with $I_i=\sigma_i/2$, $\omega_L$ is the Larmor precession frequency of nuclear spins for an external magnetic field along the $z$ direction.
We can tune the degree of non-commutativity between $B$ and $H_e$ in Eq. \eqref{Hamilt} by varying the relative strength of $\omega_L$ and $A$  [see Fig. \ref{2simul1nuclRamsey}{\color{blue}(a-b)}]:

(1) $[B,H_e]= 0$ for zero magnetic field ($\omega_L=0$). In this case, as indicated by Eq. \eqref{QChannelCAsympt_inf}, in the asymptotic limit of large $m$, sequential RIMs produce a projective measurement on the bath spin. Consequently, bath spin can be gradually polarized to $\ket{\uparrow}$ or $\ket{\downarrow}$ (eigenstates of $B$). To quantify the degree of polarization, we employ the fidelity function $\mathrm{F}(\rho_{\mathrm{fix}},\rho)=\Tr\left(\sqrt{\sqrt{\rho}\ \rho_{\mathrm{fix}}\ \sqrt{\rho}}\right)$, which describes the similarity between a given state $\rho$ and a fixed point $\rho_{\mathrm{fix}}$ that corresponds to a polarized state [see the first column in Fig. \ref{2simul1nuclRamsey}{\color{blue}(c)}].

To average trajectories leading to different fixed points separately, we divide all trajectories into distinct classes according to peak distribution in qubit measurement statistics with $X = (f_{1}-f_{0})/2=f_{1}-1/2$ [see the first column of Fig. \ref{2simul1nuclRamsey}{\color{blue}(d-f)}]. For example, $X\in [-0.5,0)$ and $X\in [0,0.5]$ label two classes of trajectories leading to $\rho_{\uparrow}=\ketbra{\uparrow}{\uparrow}{}$ and $\rho_{\downarrow}=\ketbra{\downarrow}{\downarrow}{}$ respectively. Additionally, the central locations of peaks are determined by $\expect{X}_*$ according to Eq. \eqref{f1fixed}. Particularly, when $[B,H_e]= 0$, it reduces to Eq. \eqref{f1fixedc} and reads $\expect{X_{\pm}}_*=\cos({\pm}At+\Delta \phi)/2$ with $+(-)$ for $\rho_{\uparrow}(\rho_{\downarrow})$. As $m$ increases, the broadening of each peak becomes narrower (see Appendix \ref{var0}). Note that, with fixed $\Delta \phi$, by tuning free evolution time $t$, different measurement strength can be tuned accordingly. Specifically, greater $\expect{X_{\pm}}_*$ leads to stronger measurement and faster polarization. Here we mainly show the environment steering realized by sequential weak measurements.

(2) $[B,H_e] \neq 0$ for moderate external magnetic fields comparable to the hyperfine field ($\omega_L \sim A$). The non-commutativity of $B$ and $H_e$ is noticeable and the fixed point of the channel is the maximally mixed state $\mathbb{I}/2$. Hence, as given in Eq. \eqref{ChannelQAsympt_inf}, the bath is gradually depolarized to $\mathbb{I}/2$ as $m$ increases [see the third column of Fig. \ref{2simul1nuclRamsey}{\color{blue}(c-f)}, where the only peak corresponds to the only fixed point $\mathbb{I}/2$].

(3) $[B,H_e]\approx 0$ for either weak magnetic fields ($\omega_L\ll A$) or strong magnetic fields ($\omega_L\gg A$). The non-commutativity of $B$ and $H_e$ is not significant and metastable polarization can happen. When $m$ is small, similar to the zero-field case, sequential RIMs approximately produce a projective measurement, leading to the polarization to $\ket{\uparrow}$ or $\ket{\downarrow}$. However, as $m$ increases, the fidelity starts to drop and two peaks gradually become indistinguishable, indicating the metastable polarization cannot persist over the long term and the bath finally relaxes to the maximally mixed state [see the second column of Fig. \ref{2simul1nuclRamsey}{\color{blue}(c-f)} for weak-field case where $\omega_L\ll A$].

 \begin{figure}[tp]
    \centering
    \includegraphics[width=1\columnwidth]{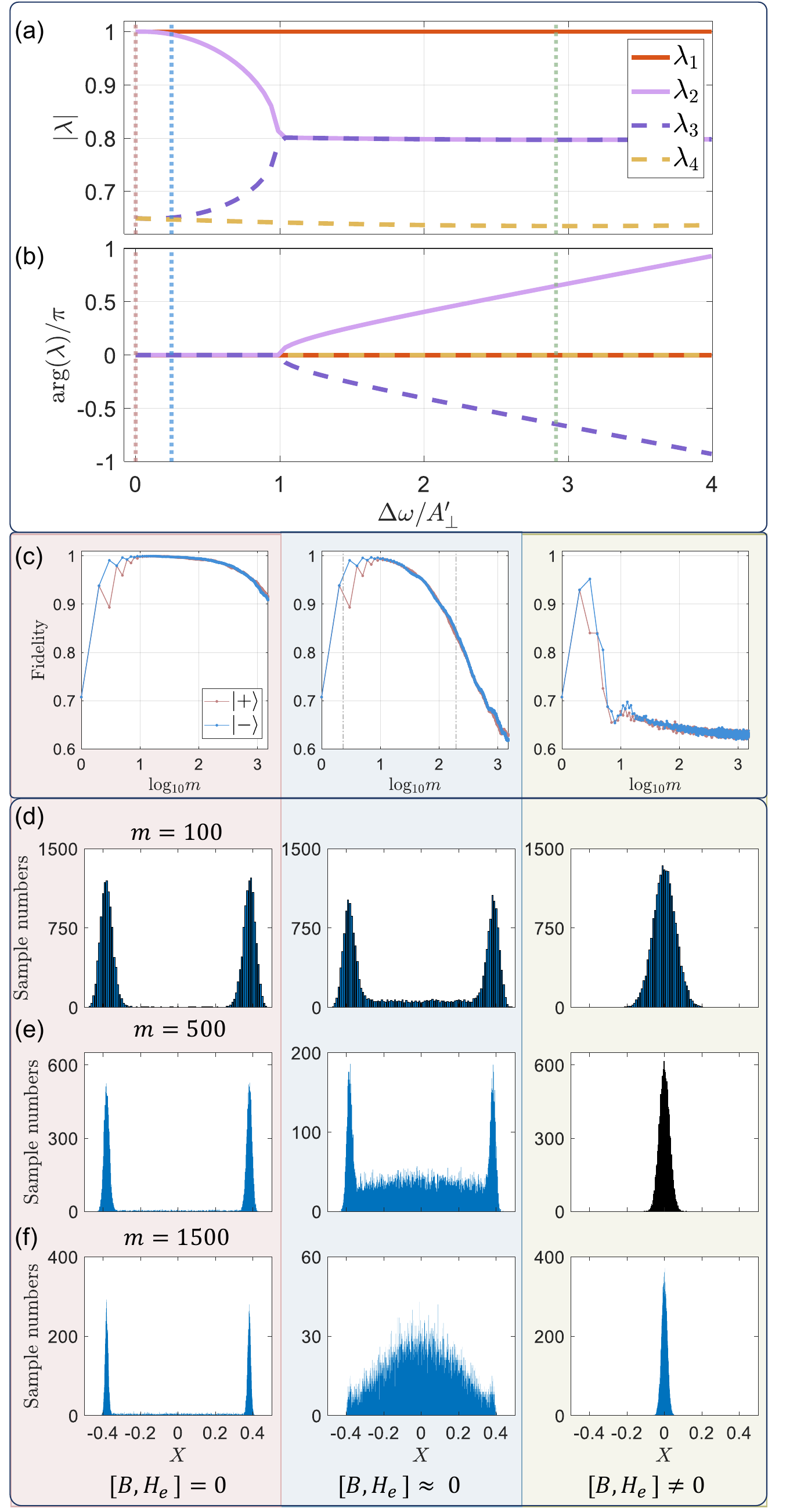}
    \caption{Monte Carlo simulations of a single-spin bath steering via sequential DD at a strong field [Eq. \eqref{HamiltDD}]. (a) Modulus and (b) argument of eigenvalues of the quantum channel when varying detuning between DD frequency and the Larmor frequency of the bath spin. For different detuning ($\Delta \omega/A_{\perp}^{\prime}=0,\ 0.25, \ 2.92$ with $A_{\perp}'=\frac{2}{\pi}A_\perp$, labeled by red, blue and green respectively), polarization, metastable polarization and depolarization can be observed. For each effect, we show (c) fidelity and (d-f) qubit measurement statistics with different number of repetitions $m$. Here trajectories with $X\in [-0.5,0)$ ($X\in [0,0.5]$) are regarded as projections to $\ket{+}$ ($\ket{-}$), which are eigenstates of $B'$. When $[B',H_e']=0$, by tuning the number $N$ of $\pi$ pulses, the strength of measurement, manifested in the locations of peaks $\expect{X_{\pm}}_* = \cos(\pm\frac{2}{\pi} A_\perp \cdot 4\tau\cdot \frac{N}{2}+\Delta \phi)/2$, can be tuned accordingly. Here we use $A=37.7$ kHz, $\tau=0.47\  \mu$s, $N=8$, $\Delta \phi=\frac{\pi}{2}$.}
    \label{3simul1nuclDD}
 \end{figure}

\subsubsection{DD sequence}
RIM sequence is a special case of DD sequence \cite{Ryan2010}, which is a commonly-used method to measure qubit coherence while decoupling certain noises. The DD sequence consists of $N$ $\pi$-pulses of the central spin inserted into free evolution between two $\frac{\pi}{2}$ pulses. For a central spin subjected to DD sequences, the noise operator and free bath Hamiltonian becomes
\begin{equation}
  B=f(t)\bm{A}\cdot\bm{I}, \quad H_e=\omega_L I_z,
\end{equation}
where $f(t)$ is a modulation function accounting for possible DD control of the qubit. Specifically, the $N$-pulse Carr-Purcell-Meiboom-Gill (CPMG) control has the form $\frac{\pi}{2}-(\tau-\pi-2 \tau-\pi-\tau)^{\frac{N}{2}}-\frac{\pi}{2}$, where a unit with period $T=4 \tau=2\pi/\omega_T$ is repeated $\frac{N}{2}$ times, so $f(t)=f(t+T)$.

To deal with time-dependent operator, we can first decompose $f(t)$ into Fourier series as $f(t)=\sum_{n=0}^{\infty} C_n \cos(n\omega_T t)$, where $C_n$ is the $n$th-order Fourier expansion coefficient. Furthermore, due to $A \ll \omega_T$, the difference between different order components is much larger than hyperfine coupling strength, and thus each component can be analyzed separately \cite{Ma2016}. We then focus on the first-order component $f(t)\approx C_0+ C_1\cos(\omega_T t)=\frac{4}{\pi}\cos(\omega_T t)$. In the weak coupling regime where $A \ll \omega_{L}$, moving to the rotating frame with respect to $\omega_T I_{z}$, adopting the rotating-wave approximation, the Hamiltonian is simplified as:
\begin{equation}
    \label{HamiltDD}
    B' = \frac{2}{\pi} A_\perp I_\perp, \ \
    H'_e = \Delta_{\omega} I_{z},
\end{equation}
where $A_\perp=\sqrt{(A_x)^2+(A_y)^2}$, $I_\perp=\cos\xi I_x+\sin\xi I_y$ with $\xi=\arctan(A_y/A_x)$, and $\Delta_{\omega} = \omega_L-\omega_T$ is the detuning between the DD frequency and the Larmor frequency of the bath spin.

Then we can tune the DD frequency to vary the effective magnetic field experienced by bath spin, and in turn change the structure of noise operator and free bath Hamiltonian:

(1) $[B',H'_e]=0$ for resonant DD ($\Delta_{\omega}=0$). In this case, the effective free bath Hamiltonian $H'_e$ disappears, which is similar to the RIM case with zero external field. Then sequential DD sequences gradually polarize the bath spin [see the first column of Fig. \ref{3simul1nuclDD}{\color{blue}(c-f)}]. Likewise, as given in Eq. \eqref{f1fixed} and Eq. \eqref{f1fixedc}, the locations of peaks are determined by eigenvalues of $B'$ in Eq. \eqref{HamiltDD} as $\expect{X_{\pm}}_* = \cos(\pm\frac{2}{\pi} A_\perp \cdot 4\tau\cdot \frac{N}{2}+\Delta \phi)/2$, with $+(-)$ for $\ket{+}$ ($\ket{-}$) which are eigenstates of $B'$ respectively. It implies, with resonant $\tau$ and fixed $\Delta \phi$, by tuning the number $N$ of $\pi$ pulses, polarization of a bath spin with different strength can be realized \footnote{Strictly speaking, $[B',H_e']$ does not accurately vanish, as we made several approximations in the derivation of effective Hamiltonian for DD sequence [Eq. \eqref{HamiltDD}]. Thus, the polarization shown in the first column of Fig. \ref{3simul1nuclDD} should be treated as metastable polarization. This accounts for the decaying of fidelity in Fig. \ref{3simul1nuclDD}(c), in contrast to that of Fig. \ref{2simul1nuclRamsey}(c) where the fidelity will not drop once reaching the high plateau.}.

(2) $[B',H'_e]\neq 0$ for off-resonant DD ($\Delta_\omega\sim A_{\perp}$). The relatively large frequency detuning makes the non-commutativity of $B'$ and $H'_e$ noticeable and induces depolarization of the bath spin, which is similar to RIM case with moderate external magnetic field [see the third column of Fig. \ref{3simul1nuclDD}{\color{blue}(c-f)}].

(3) $[B',H'_e]\approx 0$ for nearly resonant DD ($\Delta_\omega\ll A_{\perp}$). With a small frequency detuning, the non-commutativity of $B'$ and $H'_e$ is not significant and thus metastable polarization can happen, which is similar to the RIM case with weak external magnetic fields [see the second column of Fig. \ref{3simul1nuclDD}{\color{blue}(c-f)}].

\subsection{Multi-spin bath}
Now we consider a multi-spin bath. We will start with a non-interacting bath and then take account of interactions within the bath.

\subsubsection{Non-interacting multi-spin bath}
\begin{figure}[htbp]
    \centering
    \includegraphics[width=1\columnwidth]{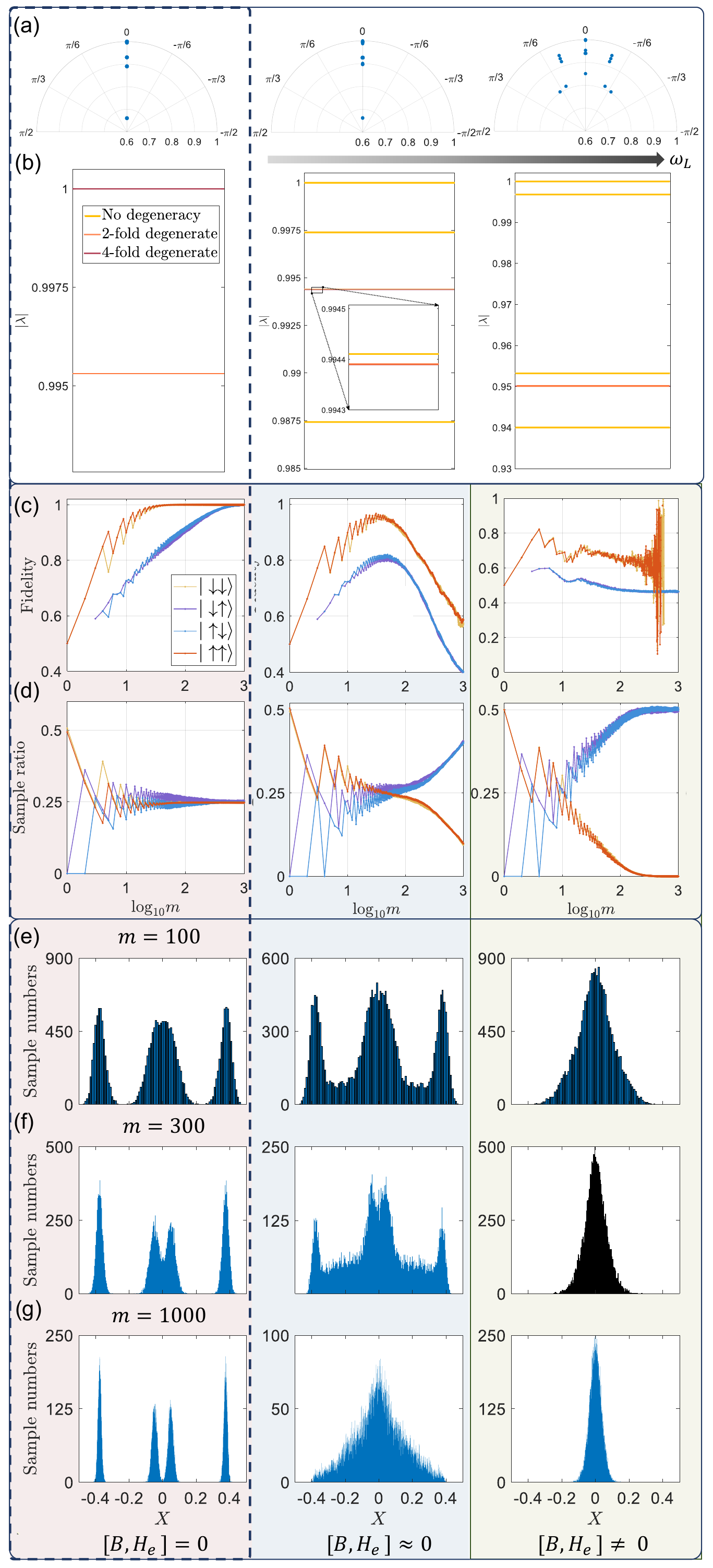}
    \caption{Monte Carlo simulations of a spin pair bath ($K=2$) steering via sequential RIMs at weak and moderate fields. The first column is for the spin pair without interaction at zero field $\omega_L=0$ [Eq. \eqref{noninter}] as a comparison, and the second and third columns are for the spin pair with interaction [Eq. \eqref{inter}]. (a) Eigenvalues on the complex plane and (b) modulus of top six largest eigenvalues of the quantum channel for the spin pair when varying external magnetic field. For different Larmor frequencies ($\omega_L=0, \ 26.8$ kHz), metastable polarization and depolarization can be observed. For each case, we show (c) fidelity, (d) sample ratio, and (e-g) qubit measurement statistics with different number of repetitions $m$. Here trajectories are divided into four classes $X\in [-0.5,-0.2)$, $[-0.2,0)$, $[0,0.2)$, $[0.2,0.5]$, corresponding to projections to $\ket{\uparrow \uparrow}$,$\ket{\uparrow \downarrow}$,$\ket{\downarrow \uparrow}$,$\ket{\downarrow \downarrow}$, respectively. Here we use $A_1=37.7$ kHz, $A_2=29.9$ kHz, $D_{12}=4.1$ kHz, $t=2\  \mu s$.}
    \label{4simul2nuclinter}
 \end{figure}

The discussions in the above subsection for a single-spin bath can be directly generalized to a bath containing $K$ independent nuclear spins, with the noise operator and bath Hamiltonian
\begin{equation}\label{noninter}
        B = \sum_{k=1}^{K}\bm{A}_k \cdot \bm{I}_k, \ H_e=\omega_L\sum_{k=1}^{K}  I_k^z.
\end{equation}
Since these nuclear spins are non-interacting, we can analyze the steering effect of sequential RIMs on each bath spin separately according to the relative strength of $A_k$ and $\omega_L$. If $\omega_L=0$ and the coupling strengths $\{A_k\}$ are inhomogeneous, the measurement statistics show $2^K$ distribution peaks and the bath can be fully polarized if these peaks are all well separated [see the first column of Fig. \ref{4simul2nuclinter}]. For a constant $\omega_L$, the bath spins near the central spin ($A_k\gg \omega_L$) can show metastable polarization, while the bath spins farther away ($A_k\sim\omega_L$ and $A_k\ll \omega_L$) are mostly depolarized (depolarization for the latter case is often due to the indistinguishable distribution peaks for the measurement statistics).

For sequential DD sequences, the noise operator and bath Hamiltonian become $B' = \frac{2}{\pi} \sum_{k=1}^K A_k^\perp I_k^\perp$, $H'_e = \sum_{k=1}^K \Delta_{\omega} I_k^{z}$ in the rotating frame with respect to $H_e$. Then the analysis is similar to the case for sequential RIMs except that we replace $A_k$, $\omega_L$ with $A_k^{\perp}$, $\Delta_{\omega}$.

\subsubsection{Interacting multi-spin bath}

\begin{figure*}[htbp]
    \centering
    \includegraphics[width=2\columnwidth]{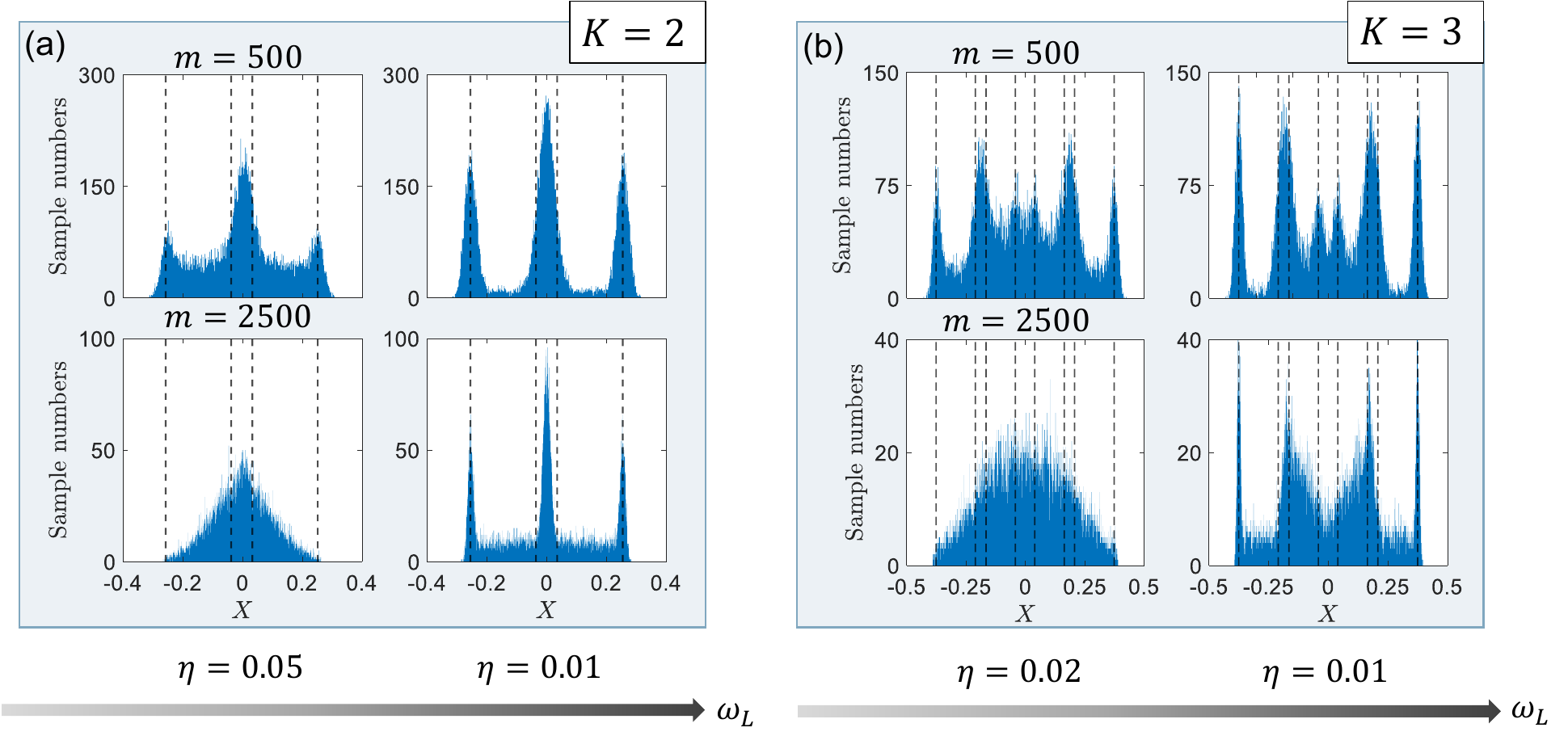}
    \caption{Monte Carlo simulations of bath steering via sequential RIMs at strong fields for interacting spin baths [Eq. \eqref{strongB}] with (a) two qubits ($K=2$) and (b) three qubits ($K=3$), respectively. Stronger field reduces the non-commutativity between $B$ and $H_e$ and can lead to longer metastable regime, where ($K+1$) peaks corresponding to ($K+1$) subspaces for an interacting $K$-spin bath can be observed. The dashed vertical lines indicate theoretical prediction of locations of peaks. Non-commutativity is quantified by $\eta=||[H_+,H_-]||/(||H_+||\cdot||H_-||)$ with $H_{\pm}=H_e\pm B$, where $||A||=\sqrt{\Tr(A^\dagger A)}$ is the Frobenius norm of a matrix. Note for the single-qubit example shown in Fig. \ref{1Qbath}{\color{blue}(d)}, $\eta=2\sqrt{2}\gamma/(1+\gamma^2)$. (a) Parameters of interacting spin pair are the same as Fig. \ref{4simul2nuclinter}, with $\omega_L t=3\pi, 9\pi$ for two different external magnetic fields. (b) For the interacting three-spin bath, we use $A_1=24.45$ kHz, $A_2=22.28$ kHz, $A_3=21.67,$ kHz, $D_{12}=0.95$ kHz, $D_{13}=0.33$ kHz, $D_{23}=0.86$ kHz, with $\omega_L t=7\pi, 17\pi$ for two different external magnetic fields.}

    \label{5simulMorenuclinter}
 \end{figure*}

We now take dipolar interactions among bath spins into account. For a $K$-spin interacting bath,
\begin{equation}\label{inter}
  B = \sum_{k=1}^{K}\bm{A}_k \cdot \bm{I}_k, \ H_e=\omega_L\sum_{k=1}^K I_k^z+\sum_{j<k} \boldsymbol{I}_j \cdot \mathbb{D}_{jk} \cdot \boldsymbol{I}_k,
\end{equation}
where $\mathbb{D}_{jk}=D_{jk}(1-3\bm{r}^T_{jk}\bm{r}_{jk}/r_{jk}^2)$ is the dipolar coupling tensor between the $j$th and $k$th bath spin with $D_{jk}=\mu_0 \gamma_{n}^2/(4\pi r_{jk}^3)$, $\vb*{r}_{jk}=[r_{jk}^x, r_{jk}^y, r_{jk}^z]$ being the displacement row vector from the $j$th spin to the $k$th spin, $\mu_0$ being the vacuum permeability, $\gamma_n$ being the gyromagnetic ratio of nuclear spins.

For zero or very weak external magnetic fields ($\omega_L\ll D_{jk}$), the channel induced by a RIM often has a single fixed point $\mathbb{I}/2^K$ and $2^K-1$ metastable points, so sequential RIMs can still cause metastable polarization of the strongly coupled bath spins ($A_k\gg \omega_L, D_{jk}$) [see the second column of Fig. \ref{4simul2nuclinter}]. While similar to the non-interacting bath, at a moderate magnetic field ($A_k\sim\omega_L$), an interacting bath can be depolarized [see the third column of Fig. \ref{4simul2nuclinter}], and the fluctuation in fidelity Fig. \ref{4simul2nuclinter}{\color{blue}(c)} results from low sample ratio in Fig. \ref{4simul2nuclinter}{\color{blue}(d)}.

With a strong external magnetic field ($\omega_L\gg A_k, D_{jk}$), the bath Hamiltonian is dominated by those energy-preserving terms,
\begin{equation}\label{strongB}
  H_e\approx \omega_L\sum_{k=1}^K I_k^z+\sum_{j<k}\frac{D_{jk}}{2}(I_j^{+}I_k^{-}+I_j^{-}I_k^{+}-4I_j^zI_k^z)
\end{equation}
with $I_k^{\pm} = I_k^x \pm i I_k^y$. Since the total spin $I_{\rm{tot}}^z=\sum_{k=1}^K I_k^z$ commutes with $H_e$, $I_{\rm{tot}}^z$ is a conserved quantity. Then the Hilbert space of bath can be decomposed into $(K+1)$ subspaces $\hilb{H}=\bigoplus_{l=-K/2}^{K/2}\hilb{H}_l $ according to $(K+1)$ components of total spin \cite{Moudgalya2022}. The dimension of each subspace $\hilb{H}_l$ is determined by $d_l=\tbinom{K}{K/2-l}=\frac{K!}{(K/2-l)!(K/2+l)!}$, corresponding to the degeneracy of each component of total spin $I_{\rm{tot}}^z=l$. Accordingly, there are $(K+1)$ fixed points, each represented by a rank-$d_l$ projector to the corresponding subspace $\hilb{H}_l$, and the measurement statistics can display $(K+1)$ peaks. However, in the presence of hyperfine interactions with central spin, represented by noise operator $B$ in Eq. \eqref{inter}, transitions among these subspaces can occur, as hyperfine difference $\bm{A}_j-\bm{A}_k$ breaks the conservation of total spin. Then metastable polarization can happen for those bath spins ($|\bm{A}_j-\bm{A}_k|\gg D_{jk}$). If magnetic field strength is increased further, the metastable regime can be extended [see Fig. \ref{5simulMorenuclinter}{\color{blue}(a)} (Fig. \ref{5simulMorenuclinter}{\color{blue}(b)}) for interacting baths with two (three) qubits, respectively].

Here we consider a small spin bath and and choose the maximally mixed state of the whole Hilbert space as the initial bath state. The main feature of non-i.i.d. statistics is the emergence of multiple distribution peaks. However, for a relatively large spin bath with a large $K$, the density of peaks become very high, and therefore different neighboring peaks may become indistinguishable for a finite number of repetitions. In this case, it is possible that the summation of all overlapping distribution peaks is reduced to the classical case with i.i.d. statistics. Moreover, the probability of finding final state in each fixed point subspace [$c_j$ in Eq. \eqref{stat}] is proportional to the dimension of each subspace. Thus, higher-dimensional subspaces $(d_l>1)$, e.g., the subspaces with $I_{\rm{tot}}^z=l$ close to $0$ with much higher degeneracy, have greater probability to be detected.

 \subsection{Noisy spin bath}
  \begin{figure*}[htbp]
 \centering
   \includegraphics[width=2\columnwidth]{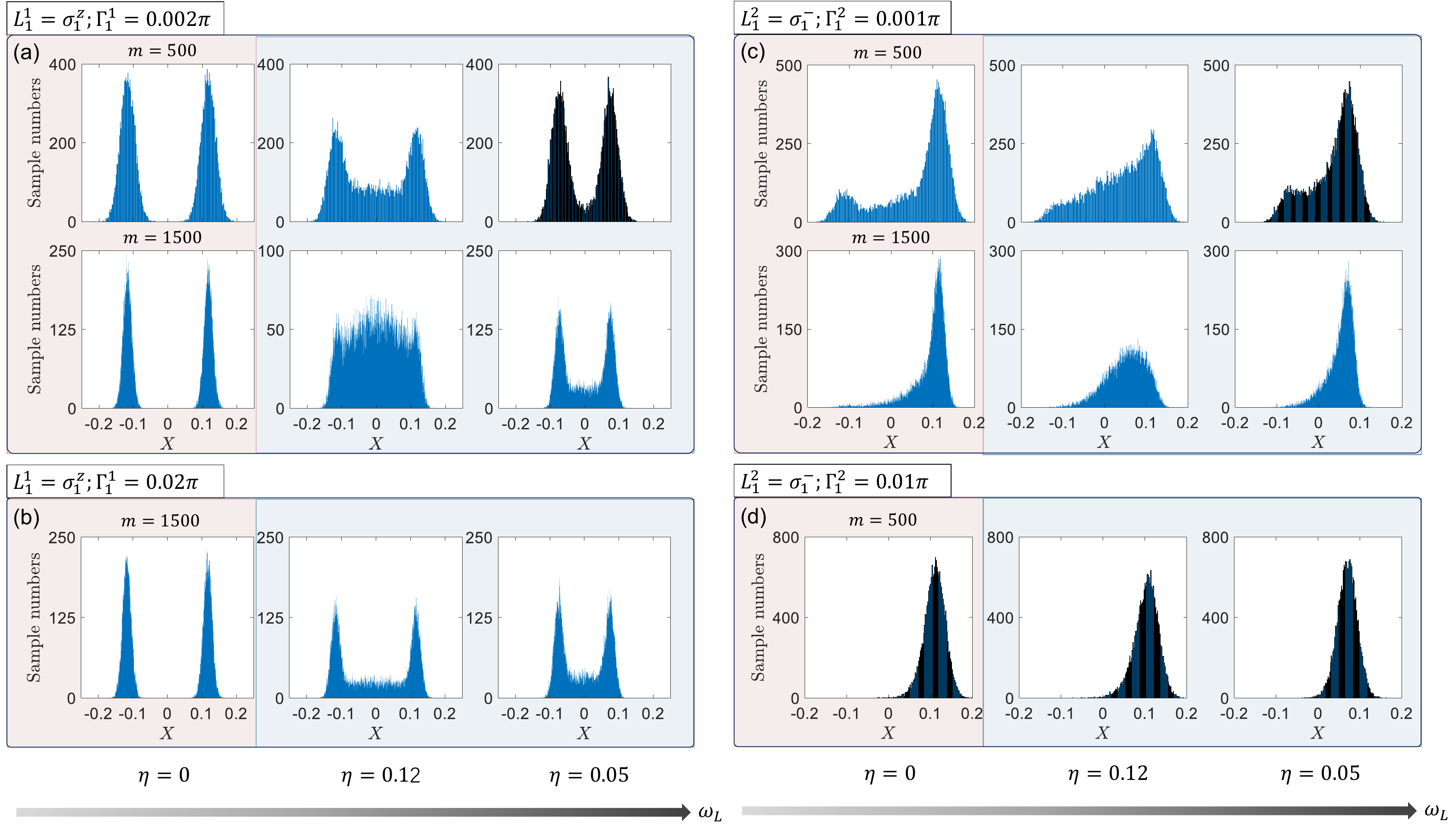}
   \caption{Monte Carlo simulations for a single-qubit bath steering via sequential RIMs in the presence of (a-b) dephasing noise ($\sigma_k^z$) and (c-d) relaxation noise ($\sigma_k^-$) respectively, with the same parameters used in Fig. \ref{2simul1nuclRamsey} with $\omega_L t=0\pi, \ 0.004\pi, \ 3\pi$ for three different external magnetic fields. $\Gamma_k^{1}$ ($\Gamma_k^{2}$) are noise strengths for dephasing (relaxation) noises on the $k$th bath spin respectively. (a-b) The polarization (first column) is unaffected by dephasing noise, while the metastable polarization at both weak and strong fields (second and third columns) regime can be extended by stronger dephasing noise. (c-d) However, relaxation noise can cause peak shift in qubit measurement statistics for both polarization and metastable polarization.}
   \label{diss1}
 \end{figure*}

 \begin{figure*}[htbp]
    \centering
      \includegraphics[width=2\columnwidth]{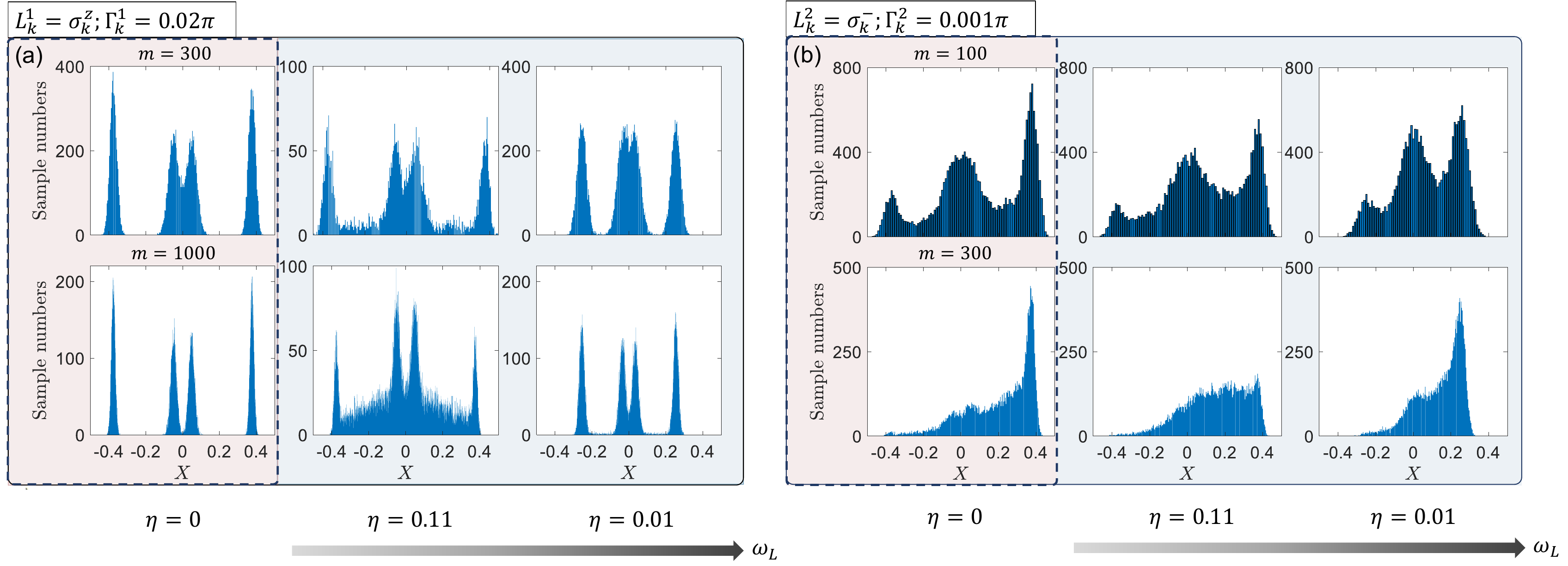}
      \caption{Monte Carlo simulations for a two-qubit bath (K=2) steering via sequential RIMs in the presence of (a) dephasing noise ($\sigma_k^z$) and (b) relaxation noise ($\sigma_k^-$) respectively, with the same parameters used in Fig. \ref{4simul2nuclinter} with $\omega_L t=0\pi, \ 21\pi$ for two different external magnetic fields. The first column is for the spin pair without interaction at zero field $\omega_L=0$ as a comparison, and the second (third) column are for the spin pair with interaction at zero (strong) field respectively.}
      \label{diss2}
    \end{figure*}

    \begin{figure*}[htbp]
        \centering
          \includegraphics[width=2\columnwidth]{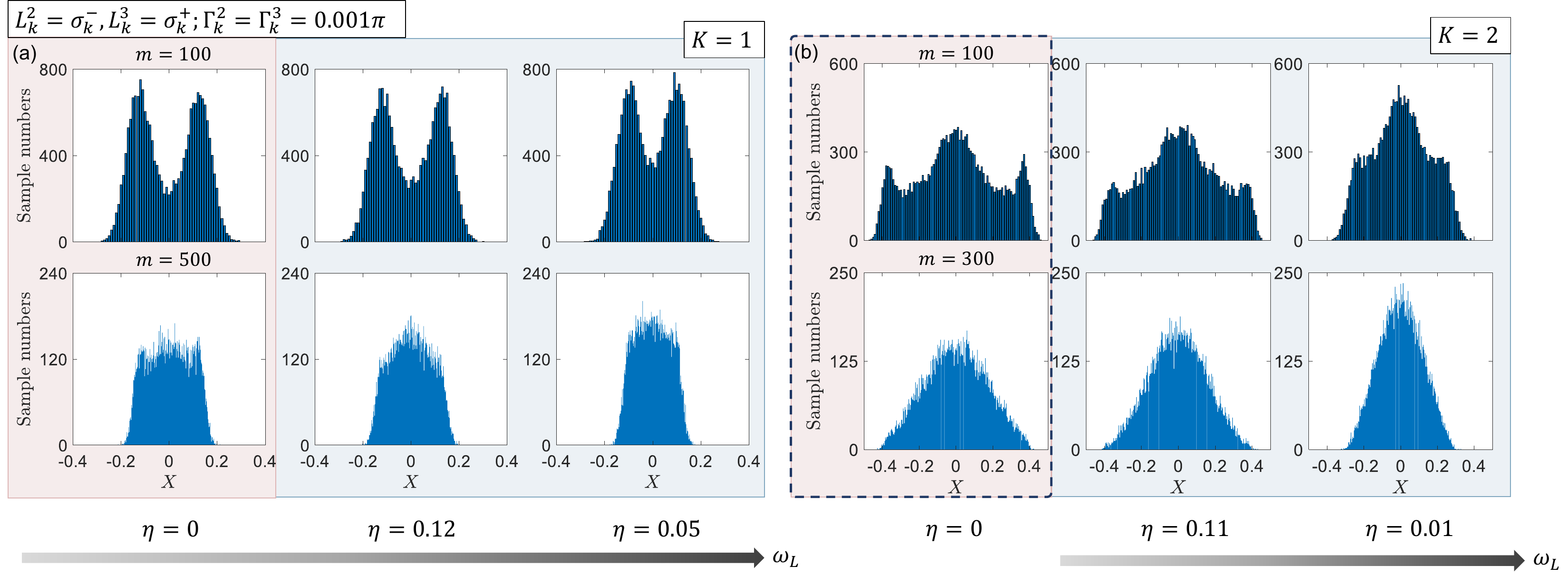}
          \caption{Monte Carlo simulations for bath steering with (a) a single spin and (b) two spins via sequential RIMs in the presence of relaxation noise ($\sigma_k^-$ and $\sigma_k^+$), with the same parameters used in Fig. \ref{diss1} and Fig. \ref{diss2}, respectively. In the presence of both jump operators with balanced noise strength, the qubit measurement statistics behave like the depolarization case.}
          \label{diss3}
        \end{figure*}
 Finally, we show that quantum polarization and metastable polarization is robust even when the environment suffers some additional noises. For simplicity, we assume that the noise of each qubit in a $K$-spin bath is independent of each other. In this case, the evolution of the composite systems $\rho_{\rm tot}$, governed by Hamiltonian given by Eq. \eqref{Hamilt}, can be described by the following Lindblad master equation,
 \begin{equation}
    \begin{aligned}
       \dv{\rho_{\rm tot}}{t}&=-i[H,\rho_{\rm tot}]\\
       &+\sum_{k=1}^K \sum_{\alpha=1}^{2^2-1}\Gamma_k^{\alpha} \left(L_k^{\alpha}\rho_{\rm tot} (L_k^{\alpha})^{\dagger}-\frac{1}{2}\left\{(L_k^{\alpha})^{\dagger} L_k^{\alpha},\rho_{\rm tot}\right\}\right),
    \end{aligned}
 \end{equation}
 where  $\{A,B\}=AB+BA$ is the anticommutator, $\Gamma_k^{\alpha}$ and $L_k^{\alpha}$ are noise strength and dissipator for the $\alpha$th type of noise on the $k$th bath spin, respectively. Here we mainly focus on the transverse dephasing error $L^1_k=\sigma^z_k$ and longitudinal relaxation errors $L^2_k=\sigma^-_k$ and $L^3_k=\sigma^+_k$. Specifically, let us take Eq. \eqref{noninter} as an example. If we further set the external magnetic field to zero, i.e., $H_e=0$, then three dissipators can be constructed by the eigenstates of $B$ for each spin as $L^1_k=\ket{\uparrow}_{k}\bra{\uparrow}-\ket{\downarrow}_{k}\bra{\downarrow}$, $L^2_k=\ket{\downarrow}_{k}\bra{\uparrow}$ and $L^3_k=\ket{\uparrow}_{k}\bra{\downarrow}$.

 We first perform Monte Carlo simulations for each type of noise separately. The effect of dephasing noise can be regarded as the rotation about the quantization axis, which mixes superposition states of eigenstates while the eigenstates remain unchanged (up to a phase). Consequently, for polarization effect where $[B,H_e]=0$, qubit measurement statistics remain unaffected in the presence of dephasing noise [see the first column of Fig. \ref{diss1}{\color{blue}(a-b)} for a single-qubit bath at zero field, and compare it with the first column of Fig. \ref{2simul1nuclRamsey}{\color{blue}(e-f)}]. Interestingly, the metastable polarization regime can be extended under stronger dephasing noise. The reason may be that the non-commutativity of $B$ and $H_e$ originating from the transverse components that mixes the eigenstates can be suppressed by dephasing noise [see the second (third) column of Fig. \ref{diss1}{\color{blue}(a-b)} for a single-qubit bath at a weak (strong) field].

 However, the jump-down error $\sigma^-_k$ of bath spin $k$ maps one eigenstate $\ket{\uparrow}_k$ to the other $\ket{\downarrow}_k$, so the peaks of measurement statistics can be shifted, for both polarization and metastable polarization [see Fig. \ref{diss1}{\color{blue}(c-d)} for a single-qubit bath]. Similar results for a two-qubit bath are shown in Fig. \ref{diss2}.
 The case with only jump-down error $\sigma^-$ is a special case when the temperature of ``true'' environment $\subt{T}{env}$ is near absolute zero, and hence the noise strength of the jump-up error $\sigma^+$ is suppressed by a Boltzmann factor $e^{-\omega \beta}$ with $\beta=1/(k_B \subt{T}{env})$ \cite{Yang2017}. For general relaxation processes with both jump-up and jump-down operators, due to the mixing of bath spin eigenstates induced by both errors, the qubit measurement statistics corresponding to the polarization case can be destroyed and behave like the depolarization one as the number of repetitions $m$ increases [see Fig. \ref{diss3}{\color{blue}(a)} (Fig. \ref{diss3}{\color{blue}(b)}) for a bath with one (two) spin(s)]. Nevertheless, for the central spin model we consider here, the spin bath usually has a memory time much longer than the duration of each RIM cycle, i.e., the noise strength is relatively low, so the steering effects can still be observed.

\section{Summary and Outlook}
We have developed a general theoretical framework to describe the steering effect on a quantum environment by sequential qubit coherence measurements, based on modeling repetitive such measurements as sequential quantum channels. By examining the structures of the noise operator and the environment Hamiltonian, we discover three distinct environment steering effects, including polarization, depolarization and metastable polarization, and elucidate the corresponding qubti measurement statistics. We have also performed extensive numerical simulations to demonstrate all kinds of steering effects by RIM and DD control sequences in a central spin model. Specifically, by varying the external magnetic field strength or frequency detuning in DD sequences, we are able to tune the non-commutativity between the noise operator and the environment Hamiltonian, and in turn tune the polarizing capability of the channel on the environment. Our rigorous analysis of the measurement statistics also clarifies what we truly measure in repetitive qubit coherence measurements.

The model discussed in our paper is applicable to a broad class of spin qubits immersed in a nuclear bath with a relatively long coherence time, such as color centers in diamond \cite{abobeih2019,Bradley2019}, quantum dots (QDs) in semiconductors \cite{Gillard2022}, single rare-earth ion qubit \cite{Ruskuc2022}, which are crucial resources for quantum information processing \cite{wolfowicz2021}. Our theory also provides a theoretical foundation for several recent experiments, in which an ancilla qubit was used to polarize a quantum environment. For example, experimental realizations of polarization to a pure state  via RIMs \cite{Maurer2012} and via DD sequence \cite{Liu2017}, as well as polarization to a subspace of Hilbert space via RIMs \cite{Dasari2022} are all special cases of our model. On the other hand, the techniques of sequential quantum channels in our model may be also applicable to the analysis of other polarization techniques using repetitive control of a central spin, e.g., repetitive dynamical nuclear polarization in GaAs QDs and NV centers \cite{Ramon2007,Jacques2009, London2013}. In a broader sense, our work also points out a potential issue in building a large-scale quantum processor, that is, characterizing the coherence of a specific qubit can seriously decrease the purity of other qubit states for most cases.

There are quite a few open problems related to the subject of our work. First, we have concentrated on the steering effect on a relatively small quantum environment (i.e., a few-spin bath), so that the qubit measurement statistics show obvious non-i.i.d. features. It will be interesting to extend the study to a medium or large quantum environment. Second, we have assumed ideal pulse control and measurement of the ancilla throughout this work, while further researches are required to study the effect of the finite-duration control pulses and the imperfect ancilla measurements, which should guide the optimization of ancilla control for improving the polarization efficiency in practical systems. It is also worthwhile to study the adaptive RIM control for further accelerating the polarization process \cite{Giedke2006,higgins2007, Scerri2020, Wang2022}. Third, it is possible to generalize the formalism in this work to study the environment steering effects induced by a more general coupling Hamiltonian between the ancilla qubit and the environment, of which an interesting example is preparation of exotic many-body states in a spin ensemble \cite{Zaporski2023,roy2020}. Finally, it will also be meaningful to consider the evironment steering effect caused by a more general ancilla, such as a multi-dimensional qudit or an infinite-dimensional bosonic mode \cite{Blais2020,Ma2021}.

\begin{acknowledgments}
    We thank Cristian Bonato and Shane Kelly for helpful discussions. C.D.Q, Y.D.J and W.L.M acknowledge support from the National Natural Science Foundation of China (Grants No. 12174379 and No. E31Q02BG), the Chinese Academy of Sciences (Grants No. E0SEBB11 and No. E27RBB11), the Innovation Program for Quantum Science and Technology (Grant No. 2021ZD0302300) and Chinese Academy of Sciences Project for Young Scientists in Basic Research (Grant No. YSBR-090). J.X.Z and G.Q.L are supported by the National Natural Science Foundation of China (Grants No. 11934018 and No. T2121001) and Chinese Academy of Sciences Project for Young Scientists in Basic Research (Grant No. YSBR-100).

\end{acknowledgments}

\appendix

\section{Decomposition of a quantum channel and the asymptotic subspace}\label{decomp}
The quantum channel has four different representations \cite{watrous2018}: the Kraus representation, the Stinespring representation, the natural representation, and the Choi representation. In this paper, we use the first three representations of quantum channels (see Appendix A of \cite{Jin2024} for a brief introduction).

As a square matrix, the natural representation of a quantum channel on HS space can be decomposed into a direct sum of Jordan blocks \cite{wolf2011url}. Precisely, with some proper invertible $d^2\times d^2$ matrix ${S}$, we have
\begin{equation}\label{Jordan}
  \begin{aligned}
	  \phii&=S\left(\bigoplus_{k=1}^\kappa \mathcal{J}_{d_k}(\lambda_k)\right)S^{-1}	\\
	  &=S\left(\sum_{\left|{\lambda}_j\right|=1} {\lambda}_j {\mathcal{P}}_j+ \sum_{\left|{\lambda}_k\right|<1} ({\lambda}_k{\mathcal{P}}_k+{\mathcal{N}}_k)\right)S^{-1} \\
      &=\bigoplus_{k=1}^\kappa \hat{\mathcal{J}}_{d_k}(\lambda_k)\\
      &=\sum_{\left|{\lambda}_j\right|=1} {\lambda}_j {\hat{\mathcal{P}}}_j+ \sum_{\left|{\lambda}_k\right|<1} ({\lambda}_k\hat{{\mathcal{P}}}_k+\hat{{\mathcal{N}}}_k),
\end{aligned}
\end{equation}
where $\mathcal{J}_{d_i}(\lambda_i)$ represents a $d_i$-dimensional Jordan block corresponding to the eigenvalue $\lambda_i$, ${\mathcal{P}}_j$ is a projection operator, ${\mathcal{N}}_k$ is a nilpotent operator satisfying ${\mathcal{N}}_{k}^{d_k}=0$ and $\hat{\mathcal{J}}_{d_k}=S{\mathcal{J}}_{d_k}S^{-1}$, $\hat{\mathcal{P}}_j=S\mathcal{P}_jS^{-1}$, $\hat{\mathcal{N}}_j=S\mathcal{N}_jS^{-1}$. Note that the Jordan blocks corresponding to the fixed points or rotating points (with eigenvalues $\abs{\lambda_j}=1$) are all rank-one projectors \footnote{See proposition 6.2 in \cite{wolf2011url}}.

We can decompose the channel into two parts as
\begin{equation}\label{decompJordan}
    \hat{\Phi} = \supoper{P} + \subt{\hat{\Phi}}{D},
\end{equation}
with
\begin{equation}
  \begin{aligned}
	  \supoper{P}&=\sum_{\lambda_j=1} \hat{\mathcal{P}}_j, \\
      \subt{\hat{\Phi}}{D}&=\sum_{\lambda_j=e^{i\phi}} {\lambda}_j {\hat{\mathcal{P}}}_j+\sum_{|{\lambda}_k|<1} ({\lambda}_k\hat{{\mathcal{P}}}_k+\hat{{\mathcal{N}}}_k),
\end{aligned}
\end{equation}
where $\hat{\mathcal{P}}$ denotes the projection to the HS space spanned by the fixed points, and $\subt{\hat{\Phi}}{D}$ denotes the projection to the HS subspace spanned by the rotating points and decaying points.
Considering that $\sum_{k=1}^\kappa d_k=d$, so if $\kappa=d$ the channel is in a diagonalizable form,
\begin{equation}\label{fixorth}
\begin{aligned}
	    \hat{\Phi}=\sum_i \lambda_i \ketHS{R_i}\braHS{L_i}, \ \mathrm{with} \ \braketHS{L_i}{R_j}{}=\delta_{ij}
\end{aligned}
\end{equation}
where $\ketHS{R_i}$ ($\ketHS{L_i}$) are right (left) eigenvectors respectively with biorthogonality, satisfying $\hat{\Phi}\ketHS{R_i}=\lambda_i\ketHS{R_i}$, $\hat{\Phi}^{\dagger}\ketHS{L_i}=\lambda_i^*\ketHS{L_i}$.

When considering sequential quantum channels, we should take a closer look at the the HS subspace spanned by the fixed points and rotating points, called \emph{asymptotic subspace} (also known as peripheral or attractor subspace). As the number of measurements $m$ increases, the projection to the decaying points ($|\lambda|<1 $) gradually vanishes due to $\lim_{m \rightarrow \infty} \lambda_i^m \rightarrow 0$ (the nilpotent part vanishes for $m\geq d_i$). Conversely, for those eigenvalues lying on the periphery of the unit disc $|\lambda|=1 $, their Jordan blocks are all one-dimensional, and they either remain unchanged or only acquire a phase during repetition. Thus, projection on the asymptotic subspace would gradually become dominant, and the quantum channel on asymptotic subspace can be represented solely by projectors
\begin{equation}
    {\hat{\Phi}}^m=\supoper{P}^m=\sum_{|\lambda_j|=1} \lambda_j^m \ketHS{R_j}\braHS{L_j}.
\end{equation}

If the channel has no rotating points, we can have the asymptotic limit
\begin{equation}
    \lim_{m \rightarrow \infty}{\hat{\Phi}}^m=\supoper{P}= \sum_{\lambda_j=1} \ketHS{R_j}\braHS{L_j}.
\end{equation}
So the behaviors of sequential channels are solely determined by the fixed points $\{R_j\}$ with $\lambda_j=1$. For the HS subspace spanned by the fixed points, we can always find a basis $\{\rho_{\rm fix}^j\}$, which are all positive operators with unit trace [Tr$(\rho_{\rm fix}^i)$=1 for any $i$] and orthogonal supports ($\rho_{\rm fix}^i\rho_{\rm fix}^j=0$ if $i\neq j$) \cite{Burgarth2013}. The set of left eigenvectors corresponding to $\{\rho_{\rm fix}^j\}$ are a set of observables $\{P_{\rm fix}^j\}$, satisfying $\braketHS{P^{j}_{\mathrm{fix}}}{\rho^{j}_{\mathrm{fix}}}{}=\delta_{ij}$. The problem now is to determine the exact form of the fixed points $\{\rho_{\rm fix}^j\}$ and the observables $\{P_{\rm fix}^j\}$ for the channel induced by a RIM.

\section{Fixed points of the channel induced by RIMs}\label{Proposition}
The fixed points $\{\rho_{\rm fix}^j\}$ of the channel ${\hat{\Phi}}$ induced by RIMs have been analyzed in \cite{Jin2024} to study the metastability phenomena in sequential RIMs. For completeness of the presentation, below we first reproduce the results in \cite{Jin2024}.

\textbf{Proposition 1}. The fixed points of the channel in Eq. (\ref{RIMKraus}) depend on the commutativity of $B$ and $H_e$. If $[B,H_e]=0$, the fixed points are spanned by a set of rank-one projections $\{|j\rangle\langle j|\}_{j=1}^d$; if $[B,H_e]\neq 0$, the fixed points are spanned by a set of projection operators $\{\Pi_j\}_{j=1}^r$ ($r\leq d$), satisfying $\sum_{j=1}^r\Pi_j=\mathbb{I}$.

\begin{proof}
It has been proven that $\rho$ is a fixed point of a unital channel if and only if it commutes with every Kraus operator (see Theorem 4.25 in \cite{watrous2018}), i.e., $[\rho, M_{\alpha}]=0$ for any $\alpha$. This implies that $[\rho, U_0]=[\rho,U_1]=0$. If the above condition is always satisfied for any $\alpha$, then $[\rho,B]=[\rho,H_e]=0$.

If $[B,H_e]=0$, then $B$ and $H_e$ can be diagonalized simultaneously, $B=\sum_{j=1}^d b_j \ket{j}\bra{j}$ and $H_e=\sum_{j=1}^d \varepsilon_j \ket{j}\bra{j}$. So the fixed points must include the rank-one projections $\{|j\rangle\langle j|\}_{j=1}^d$ and their linear combinations.

If $[B,H_e]\neq 0$, we can block diagonalize them simultaneously by unitary transformation,
\begin{equation}
  B=W\left(\bigoplus_{j=1}^r B_j\right) W^{\dagger}, \quad H_e=W\left(\bigoplus_{j=1}^r H_j\right) W^{\dagger}
\end{equation}
where $r\leq d$ is the number of blocks (with equality occurring only when $[B,H_e]=0$ and all of blocks are one-dimensional), $W$ is unitary matrix and should be chosen so that $B_j$ and $H_j$ for any $j$ cannot be reduced further to have more blocks. There must be at least one subspace $\h_j$ in which $[B_j,H_j]\neq 0$ to make $[B,H_e]\neq 0$. Such a block diagonalization partitions the Hilbert space of the target system into the direct sum of $r$ subspaces $\mathcal{H}=\bigoplus_{j=1}^r\mathcal{H}_j$, and $[B_j, H_{j}] \neq 0$ for at least one subspace $\mathcal{H}_j$ with ${\rm dim}(\mathcal{H}_j)\geq2$. Thus, the Kraus operator is also transformed to a block-diagonal form as $M_\alpha=\bigoplus_{j=1}^r M_\alpha^j$. Then the fixed points must include the set of projections $\{\Pi_j\}_{j=1}^r$ ($r\leq d$) and their linear combinations, where $\Pi_j=WP_jW^{-1}$ with $P_j$ being the projector to $\mathcal{H}_j$. Note that the case $[B,H_e]=0$ can be regarded as a special case of $[B,H_e]\neq 0$.

Now we prove that there are no other fixed points for the case $[B,H_e]\neq 0$, where there is at least one block with $[B_j,H_j]\neq 0$ and $[M_0^j,M_1^j]\neq0$.
Suppose there is another density matrix satisfying $[\rho',M_\alpha^j]=0$. If $\rank{\rho'}=d_j$, then $[\rho',M_0^j]=[\rho',M_1^j]=0$. Since the positive operator $\rho'$ can be diagonalized, this implies that $[M_0^j,M_1^j]=0$. If $\rank(\rho')<d_j$, then formulate another fixed point $\rho''=\rho'+\eta\mathbb{I}$ with $\eta$ being a positive number such that $\rank(\rho'')=d_j$, then the proof is similar to the former case.

\end{proof}

The observables $\{P_{\rm fix}^j\}$ actually correspond to the fixed points of the dual channel ${\Phi}^{\dagger}=\sum_{\alpha=1}^r M_{\alpha}^\dagger (\cdot) M_{\alpha}$, which can easily derived from its definition $\Tr[B\Phi(A)]=\Tr[\Phi^{\dagger}(B)A]$. Since the channel is trace-preserving, so $\hat{\Phi}^{\dagger}(\mathbb{I})=\mathbb{I}$ is unital. Then we can use similar reasoning to show that the fixed points of $\hat{\Phi}^{\dagger}$ are the same as $\hat{\Phi}$. However, due to the constraint $\braketHS{P^{j}_{\mathrm{fix}}}{\rho^{j}_{\mathrm{fix}}}{}=\delta_{ij}$, $P_{\rm fix}^j$ is different from $\rho_{\rm fix}^j$ by a normalization factor.
Specifically, we have
\begin{equation}\label{projfix}
    \lim_{m \rightarrow \infty}{\hat{\Phi}}^m=\supoper{P}=\sum_j\ketbraHS{\rho^{j}_{\mathrm{fix}}}{P^{j}_{\mathrm{fix}}}{},
\end{equation}
with $P^{j}_{\mathrm{fix}}=d_j\rho^{j}_{\mathrm{fix}}$ with $d_j$ being the rank of $\rho^{j}_{\mathrm{fix}}$.

\section{Statistics of sequential qubit coherence measurements}\label{qstatistics}
For $m$ sequential coherence measurements, one obtains a set of measured data $\{\alpha_1, \cdots, \alpha_m\}$. For any $n \in m$, $\alpha_n\in\{a_r\}_{r=0}^1$, corresponding to outcomes of two Kraus operators $\{M_r\}_{r=0}^1$ respectively. We are interested in the average of all outcomes $\bar{\alpha}=\frac{1}{m}\sum_{n=1}^m \alpha_n$. If we specify $\{a_0=0,a_1=1\}$, it coincides with $f_1$.

\subsection{Expectation}\label{exp0}
With the probability given in Eq. \eqref{pniid}, its expectation is then determined by
\begin{equation}
    \begin{aligned}
     \expect{f_1} &= \sum_{\alpha_1} \cdots \sum_{\alpha_m} f_1 p(\alpha_1, \cdots, \alpha_m|\rho)\\
     &=\frac{1}{m} \sum_{\alpha_1} \cdots \sum_{\alpha_m} \sum_{n=1}^m \alpha_n\braketHS{\mathbb{I}}{ \supoper{M}_{{\alpha_m}} \cdots \supoper{M}_{\alpha_1}|\rho}{}.
    \end{aligned}
\end{equation}
In general, the Kraus operators may not commute with each other. Fortunately, there are two facts that can be used to simplify calculations
\begin{equation}\label{twofacts}
    \braHS{\mathbb{I}}\hat{\Phi}=\braHS{\mathbb{I}}\hat{\Phi}^m=\braHS{\mathbb{I}}, \ \ \sum_{\alpha_i\in\{a_0,a_1\}}^{}\supoper{M}_{\alpha_i}=\sum_{r=0}^{1}\supoper{M}_{r}=\hat{\Phi},
\end{equation}
where the first one says the identity is a left eigenvector of the channel, and the second one says the summation of operators in each measurement can be reduced to the summation of two possible Kraus operators. If we first split the summation as
\begin{equation}
    \begin{aligned}
         \expect{f_1}
         =&\frac{1}{m} \sum_{\alpha_1} \cdots \sum_{\alpha_{m-1}} \sum_{n=1}^{m-1} \alpha_n \left(\sum_{\alpha_m}p(\alpha_1, \cdots, \alpha_m|\rho)\right)\\
         +&\frac{1}{m} \sum_{\alpha_1} \cdots \sum_{\alpha_{m-1}} \sum_{\alpha_m} \alpha_m p(\alpha_1, \cdots, \alpha_m|\rho),
        \end{aligned}
\end{equation}
then in view of Eq. \eqref{twofacts}, two terms can be simplified using
\begin{equation}
    \begin{aligned}
       \sum_{\alpha_m}p(\alpha_1, \cdots, \alpha_m|\rho)&=\braketHS{\mathbb{I}}{(\sum_{\alpha_m} \supoper{M}_{{\alpha_m}}) \supoper{M}_{{\alpha_{m-1}}} \cdots \supoper{M}_{\alpha_1}|\rho}{}\\
        &=\braketHS{\mathbb{I}}{ \supoper{M}_{{\alpha_{m-1}}} \cdots \supoper{M}_{\alpha_1}|\rho}{}\\
        &=p(\alpha_1, \cdots, \alpha_{m-1}|\rho)
    \end{aligned}
\end{equation}
and
\begin{equation}
    \begin{aligned}
        & \sum_{\alpha_1} \cdots \sum_{\alpha_{m-1}} \sum_{\alpha_m} \alpha_m p(\alpha_1, \cdots, \alpha_m|\rho)\\
        &= \braketHS{\mathbb{I}}{(\sum_{\alpha_m}\alpha_m \supoper{M}_{\alpha_m})  (\sum_{\alpha_{m-1}}\supoper{M}_{{\alpha_{m-1}}}) \cdots (\sum_{\alpha_1}\supoper{M}_{\alpha_1})|\rho}{}\\
        &=\braketHS{\mathbb{I}}{(\sum_{r=0,1}\alpha_r \supoper{M}_r) \hat{\Phi}^{m-1}|\rho}{},
    \end{aligned}
\end{equation}
respectively. With these simplifications, it yields
\begin{equation}
    \begin{aligned}
         \expect{f_1}
         =&\frac{1}{m} \sum_{\alpha_1} \cdots \sum_{\alpha_{m-2}} \sum_{n=1}^{m-2} \alpha_n \left(\sum_{\alpha_{m-1}}p(\alpha_1, \cdots, \alpha_{m-1}|\rho)\right)\\
         +&\frac{1}{m} \sum_{\alpha_1} \cdots \sum_{\alpha_{m-2}} \sum_{\alpha_{m-1}} \alpha_{m-1} p(\alpha_1, \cdots, \alpha_{m-1}|\rho)\\
          + &\frac{1}{m}\braketHS{\mathbb{I}}{(\sum_{r=0,1}\alpha_r \supoper{M}_r) \hat{\Phi}^{m-1}|\rho}{}.
        \end{aligned}
\end{equation}
In this iterative way, we find
\begin{equation}\label{exp1}
    \begin{aligned}
    \expect{f_1} &=\frac{1}{m}\braketHS{\mathbb{I}}{\sum_{r=0}^{1}a_r \supoper{M}_r \sum_{n=1}^{m}\hat{\Phi}^{n-1}|\rho}{}\\
    &=\frac{1}{m}\braketHS{\mathbb{I}}{\supoper{M}_1 \sum_{n=1}^{m}\hat{\Phi}^{n-1}|\rho}{},
    \end{aligned}
\end{equation}
where in the second step we specified $\{a_0=0,a_1=1\}$.

To factor out the contributions of fixed points, we use the decomposition of quantum channel in Eq. \eqref{decompJordan} and the following relation
\begin{equation}\label{PhiDorth}
    \supoper{P}_j \subt{\hat{\Phi}}{D}=\subt{\hat{\Phi}}{D}\supoper{P}_j =0,
\end{equation}
then the summation of the channel becomes
\begin{equation}\label{phipower}
    \frac{1}{m}\sum_{n=1}^{m}\hat{\Phi}^{n-1}=\sum_{j=1}^J \supoper{P}_j+ \frac{1}{m}\sum_{n=1}^{m}\subt{\hat{\Phi}}{D}^{n-1}\supoper{Q},
\end{equation}
where $\supoper{Q}=\supoper{I}-\supoper{P}$ is the projection out of the fixed point space with $\supoper{I}$ being the identity operator on the HS space. Plugging the above equation into Eq. \eqref{exp1}, we reach the equation in Eq. \eqref{stat}
\begin{equation}
    \expect{f_1} = \sum_{j=1}^J c_j \expect{f_{1j}}_{*}+\frac{1}{m} \sum_{n=1}^{m}\langle\langle \mathbb{I} |\supoper{M}_1 \subt{\hat{\Phi}}{D}^{n-1}\supoper{Q} |\rho\rangle\rangle,
\end{equation}
where $c_j=\Tr(P^{j}_{\mathrm{fix}} \rho)$ amounts to the probability of obtaining $j$th fixed point given initial state $\rho$, and
\begin{equation}\label{exp2}
    \expect{f_{1j}}_{*}=\braketHS{\mathbb{I}}{\supoper{M}_1|\rho^{j}_{\mathrm{fix}}}{}=\sum_{r=0}^{1}a_r\braketHS{\mathbb{I}}{\supoper{M}_r|\rho^{j}_{\mathrm{fix}}}{}=\expect{\alpha_j}_*.
\end{equation}
where for the last step we used Eq. \eqref{mprob} so that it is equivalent to $\sum_{r=0}^{1}a_r p(a_r|\rho^{j}_{\mathrm{fix}})=\expect{\alpha_j}_*$, which is the expectation of a single measurement on $j$th fixed point. This is similar to the expecation obtained in a classical scenario where the measurement statistics are i.i.d., satisfying $\expect{\bar{\alpha}_j}_*=\expect{\alpha_j}_*$.

\subsection{Variation}\label{var0}
To estimate the peak broadening of measurement distribution as repetitions increase, we further examine the variation with respect to each fixed point. As discussed in the Sec. \ref{effect} in main text, for each round of $m$-fold repetitions, an arbitrary initial state of environment will be projected to the subspace of one of fixed points with probability $c_j=\Tr(P^{j}_{\mathrm{fix}} \rho)$ given in Eq. \eqref{stat}. Without loss of generality, we examine the variation obtained in the subspace of each fixed point $\rho^{j}_{\mathrm{fix}}$,
\begin{equation}
    \begin{aligned}
         \mathrm{Var}[f_{1j}]_* &= \expect{(f_{1}-\expect{f_{1j}}_*)^2}_*\\
         &=\frac{1}{m^2}\left\langle\sum_{n=1}^m [(\delta \alpha_n)_{j}]^2 + 2 \sum_{n=1}^m \sum_{q>n}^m (\delta \alpha_n)_j (\delta \alpha_q)_j \right\rangle_*,
    \end{aligned}
\end{equation}
where $(\delta \alpha_n)_j=\alpha_n-\expect{f_{1j}}_*$, we added the asterisk $*$ as subscript to indicate that the probability is obtained when the initial state is a fixed point, i.e., $p(\alpha_1,\cdots,\alpha_m|\rho^{j}_{\mathrm{fix}})$. Again, using two useful facts given in Eq. \eqref{twofacts}, the summation can be simplified as
\begin{equation}
    \begin{aligned}
      &\sum_{\alpha_1} \cdots \sum_{\alpha_n} \cdots \sum_{\alpha_q} \cdots \sum_{\alpha_m} \\
      &(\delta \alpha_n)_j (\delta \alpha_q)_j \braketHS{\mathbb{I}}{\supoper{M}_{{\alpha_m}} \cdots \supoper{M}_{{\alpha_q}} \cdots \supoper{M}_{{\alpha_n}} \cdots \supoper{M}_{\alpha_1}|\rho^{j}_{\mathrm{fix}}}{}\\
      &= \braketHS{\mathbb{I}}{\supoper{E}_j^{(1)}\hat{\Phi}^{q-n-1}\supoper{E}_j^{(1)}\hat{\Phi}^{n-1}|\rho^{j}_{\mathrm{fix}}}{},
    \end{aligned}
\end{equation}
where
\begin{equation}\label{deltaa}
    \supoper{E}_j^{(n)}=\sum_{r=0}^{1}[(\delta a_r)_j]^n \supoper{M}_r, \ (\delta a_r)_j=a_r-\expect{f_{1j}}_*.
\end{equation}
Then it turns out
\begin{equation}\label{var1fix}
    \begin{aligned}
     \mathrm{Var}[f_{1j}]_{*} &=\frac{1}{m^2}\sum_{n=1}^{m}\braketHS{\mathbb{I}}{\supoper{E}_{j}^{(2)} \hat{\Phi}^{n-1}|\rho^{j}_{\mathrm{fix}}}{}{}\\
    &+\frac{2}{m^2}\sum_{n=1}^{m-1} \sum_{q=n}^{m-1}\braketHS{\mathbb{I}}{\supoper{E}_{j}^{(1)} \hat{\Phi}^{q-n} \supoper{E}_{j}^{(1)}\hat{\Phi}^{n-1}|\rho^{j}_{\mathrm{fix}}}{}{}.
    \end{aligned}
\end{equation}
Plugging Eq. \eqref{phipower} into Eq. \eqref{var1fix}, exploiting two orthogonality given in Eq. \eqref{PhiDorth} and Eq. \eqref{fixorth} to eliminate terms containing $\subt{\hat{\Phi}}{D}\supoper{Q}\ketHS{\rho^{j}_{\mathrm{fix}}}$ and to simplify $\sum_{i=1}^J \supoper{P}_i \ketHS{\rho^{j}_{\mathrm{fix}}}=\ketHS{\rho^{j}_{\mathrm{fix}}}$ respectively, it follows
\begin{equation}\label{var2fix}
    \begin{aligned}
        \mathrm{Var}[f_{1j}]_* &=\frac{1}{m}\braketHS{\mathbb{I}}{\supoper{E}_{j}^{(2)}|\rho^{j}_{\mathrm{fix}}}{}\\
        &+\frac{2}{m^2}\sum_{n=1}^{m-1}\sum_{q=n}^{m-1}\left(\sum_{i=1}^{J} \braketHS{\mathbb{I}}{\supoper{E}_j^{(1)} \ketHS{\rho^{i}_{\mathrm{fix}}}\braHS{P^{i}_{\mathrm{fix}}} \supoper{E}_j^{(1)}|\rho^{j}_{\mathrm{fix}}}{}\right.\\
        &\ \ \ \ \ \ \ \ \ \ \ \ \ \ \ \ \ \ \ \ \ \ \ \ \left.+\braketHS{\mathbb{I}}{\supoper{E}_j^{(1)} \subt{\hat{\Phi}}{D}^{q-n}\supoper{Q}\supoper{E}_j^{(1)}|\rho^{j}_{\mathrm{fix}}}{}\right).
    \end{aligned}
\end{equation}
According to Eq. \eqref{exp2}, noting that
\begin{equation}
    \braketHS{\mathbb{I}}{\supoper{E}_{j}^{(1)}|\rho^{j}_{\mathrm{fix}}}{}=\expect{\alpha_j-\expect{f_{1j}}_{*}}_{*}=0,
\end{equation}
then the terms containing $\braHS{\mathbb{I}}{\supoper{E}_{j}^{(1)}\supoper{P}_{j}}$ vanish, and the summation $\sum_{i=1}^{J}$ should be replaced by $\sum_{i\neq j}^{J}$. However, for $i\neq j$, $\braHS{P^{i}_{\mathrm{fix}}} \supoper{M}_r\ketHS{\rho^{j}_{\mathrm{fix}}}=0$ for every $\supoper{M}_r$ due to orthogonality of fixed points, which follows $\braHS{P^{i}_{\mathrm{fix}}} \supoper{E}_j^{(1)}\ketHS{\rho^{j}_{\mathrm{fix}}}=0$ and
\begin{equation}
    \begin{aligned}
        \mathrm{Var}[f_{1j}]_* &=\frac{1}{m}\braketHS{\mathbb{I}}{\supoper{E}_{j}^{(2)}|\rho^{j}_{\mathrm{fix}}}{}\\
        &+\frac{2}{m^2}\sum_{n=1}^{m-1}\sum_{q=n}^{m-1}\braketHS{\mathbb{I}}{\supoper{E}_j^{(1)} \subt{\hat{\Phi}}{D}^{q-n}\supoper{Q}\supoper{E}_j^{(1)}|\rho^{j}_{\mathrm{fix}}}{}.
    \end{aligned}
\end{equation}
We can further expand the series using
\begin{equation}\label{phipower2}
    \sum_{n=0}^{m-1}\subt{\hat{\Phi}}{D}^{n}= \frac{\supoper{I}-\subt{\hat{\Phi}}{D}^m}{\supoper{I}-\subt{\hat{\Phi}}{D}} ,
\end{equation}
which is valid as $\subt{\hat{\Phi}}{D}$ has no unit eigenvalue ($\lambda=1$), and then there exists the inverse $(\supoper{I}-\subt{\hat{\Phi}}{D})^{-1}$ with $\supoper{I}$ being the identity. Then we obtain
\begin{equation}\label{var3fix}
    \begin{aligned}
    \mathrm{Var}[f_{1j}]_* &=\frac{1}{m}\sigma^2_{j}\\
    &-\frac{2}{m^2}\braketHS{\mathbb{I}}{\supoper{E}_j^{(1)} \frac{\subt{\hat{\Phi}}{D}-\subt{\hat{\Phi}}{D}^{m}}{(\supoper{I}-\subt{\hat{\Phi}}{D})^2}\supoper{Q}\supoper{E}_j^{(1)}|\rho^{j}_{\mathrm{fix}}}{}
    \end{aligned}
\end{equation}
with
\begin{equation}\label{vars}
    \sigma^2_{j}=\braketHS{\mathbb{I}}{\supoper{E}_{j}^{(2)}|\rho^{j}_{\mathrm{fix}}}{}+2 \braketHS{\mathbb{I}}{\supoper{E}_j^{(1)} \frac{\supoper{Q}}{\supoper{I}-\subt{\hat{\Phi}}{D}} \supoper{E}_j^{(1)}|\rho^{j}_{\mathrm{fix}}}{}.
\end{equation}
This indicates the broadening of each peak shrinks as $\frac{1}{m}$. And note the first term in Eq. \eqref{vars} is similar to the variation of an i.i.d. variable,
\begin{equation}
    \braketHS{\mathbb{I}}{\supoper{E}_{j}^{(2)}|\rho^{j}_{\mathrm{fix}}}{}=\expect{\alpha_j^2}_*-\expect{\alpha_j}_*^2=\mathrm{Var}[\alpha_j]_*.
\end{equation}

It should be noted that the above results only show that the variation scales as $\frac{1}{m}$ when the initial environment state is some fixed point of the channel, while it is still an open problem to extend this analysis to an arbitrary initial environment state, although we have numerical evidence that the above conclusion holds for this general case.

\subsection{Asymptotic limit}\label{qstatisticsLim}
In this section, we show that the expectation of measurement average is only determined by fixed points in the asymptotic limit, which means the second term in Eq. \eqref{stat} will vanish as $m\rightarrow \infty$. To deal with the series $\sum_{n=1}^{m}\subt{\hat{\Phi}}{D}^{n-1}$, we can further divide $\subt{\hat{\Phi}}{D}$ in terms of rotating points ($|\lambda_l| = 1$ but $\lambda_l \neq 1$) and decaying points ($|\lambda_k| < 1$) as in Eq. \eqref{decompJordan}. For each rotating point,
\begin{equation}
    \lim_{m \rightarrow \infty} \frac{1}{m}\sum_{n=1}^{m} e^{i n \varphi_l}=\lim_{m \rightarrow \infty} \frac{1}{m}\frac{1-e^{i m \varphi_l}}{1-e^{i \varphi_l}}=0.
\end{equation}
For the decaying points, each Jordan block $\supoper{J}_k$ with dimension $d_k$ has no eigenvalue equal to one, and then there exists $(\supoper{I}_{d_k}-\supoper{J}_k)^{-1}$, where $\supoper{I}_{d_k}$ is the identity in $d_k$-dimensional subspace. By noting that each $\supoper{J}_k^m$ is convergent due to $|\lambda_k| < 1$ \cite{horn2012matrix}, it yields
\begin{equation}
    \lim_{m \rightarrow \infty} \frac{1}{m}\sum_{n=1}^{m} \supoper{J}_k^n = \lim_{m \rightarrow \infty} \frac{1}{m} \frac{\supoper{I}_{d_k}-\supoper{J}_k^m}{\supoper{I}_{d_k}-\supoper{J}_k}=0.
\end{equation}
Therefore, as $m\rightarrow \infty$, the second term in Eq. \eqref{stat} vanishes, implying that the contributions from outside the fixed point space cannot survive, so we have
\begin{equation}
    \lim_{m \rightarrow \infty} \expect{f_1} = \sum_{j=1}^J c_j \expect{f_{1j}}_{*}.
\end{equation}

\bibliography{refer}
\newpage

\end{document}